\documentclass[preprint,pre,aps,amsfonts,amsmath,showpacs,superscriptaddress,nofootinbib]{revtex4}
\usepackage{amsmath,amssymb,amscd}
\usepackage{empheq}
\usepackage{graphicx}
\usepackage{bm}
\usepackage{color}

\newcommand{\m}{\mathfrak{m}}

\newcommand{\bea}{\begin{eqnarray}}
\newcommand{\eea}{\end{eqnarray}}

\def\XXint#1#2#3{{\setbox0=\hbox{$#1{#2#3}{\int}$}
     \vcenter{\hbox{$#2#3$}}\kern-.5\wd0}}

\begin{document}
%
%
\title{Statistics of the Number of Records for Random Walks and L\'evy Flights on a $\bm{1D}$ Lattice}
\author{Philippe Mounaix}
\email{philippe.mounaix@polytechnique.edu}
\affiliation{CPHT, CNRS, Ecole
Polytechnique, IP Paris, F-91128 Palaiseau, France.}
\author{Satya N. Majumdar}
\email{satya.majumdar@lptms.u-psud.fr}
\affiliation{LPTMS, CNRS, Univ. Paris-Sud, Universit\'e Paris-Saclay, 91405 Orsay, France.}
\author{Gr\'egory Schehr}
\email{gregory.schehr@lptms.u-psud.fr}
\affiliation{LPTMS, CNRS, Univ. Paris-Sud, Universit\'e Paris-Saclay, 91405 Orsay, France.}
\date{\today}
\begin{abstract}
We study the statistics of the number of records $R_n$ for a symmetric, $n$-step, discrete jump process on a $1D$ lattice. At a given step, the walker can jump by arbitrary lattice units drawn from a given symmetric probability distribution. This process includes, as a special case, the standard nearest neighbor lattice random walk.  We derive explicitly the generating function of the distribution $P(R_n)$ of the number of records, valid for arbitrary discrete jump distributions. As a byproduct, we provide a relatively simple proof of the generalized Sparre Andersen theorem for the survival probability of a random walk on a line, 
with discrete or continuous jump distributions. For the discrete jump process, we then derive the 
asymptotic large $n$ behavior of $P(R_n)$ as well as of the average number of records $E(R_n)$. We show that unlike the case of random walks with symmetric and continuous jump distributions where the record statistics is {\em strongly} universal (i.e., independent of the jump distribution for all $n$), the record statistics for lattice walks depends on the jump distribution for any fixed $n$. However, in the large $n$ limit, we show that the distribution of the scaled record number $R_n/E(R_n)$ approaches a universal, half-Gaussian form for any discrete jump process. The dependence on the jump distribution enters only through the scale factor $E(R_n)$, which we also compute in the large $n$ limit for arbitrary jump distributions. We present explicit results for a few examples and provide numerical checks of our analytical predictions.
\end{abstract}
\maketitle
%
%
\section{Introduction}\label{sec1}

Consider a discrete time series $\lbrace x_0,\, x_1,\, x_2,\, \cdots ,\, x_n\rbrace$ of $n$ entries representing, e.g., the temperature in a city or the depth of a river or the stock price of a company, all these data being taken on, say, a daily basis. Then a record is said to happen on day $i$ if the $i$th entry $x_i$ is larger than all previous entries $x_0,\, x_1,\, x_2,\, \cdots ,\, x_{i-1}$. In general, record statistics is expected to be of interest in fields where time series are used and where the size of the entries is a relevant parameter. Such fields include meteorology \cite{Hoyt1981,RP2006}, hydrology \cite{Matalas1997,VZM2001}, insurance and risk management, trading \cite{WBK2011,WMS2012,C2015}, economics \cite{Barlevy2002,BN2006}, sports \cite{GTS2002,GTS2007,Glick1978,Krug2007,BRV2007}, etc. Another important application concerns current issues of climatology such as global warming where both temperature records~\cite{RP2006,MTWEM2009,WK2010,AK2011,FWK2012,WHK2014}
and rainfall precipitation records~\cite{WW1999,Benestad2006,MVK2019} play an important role in anticipating future climatic conditions. Statistics of record events have also been found relevant in biology \cite{KJ2005}, in the theory of spin glasses \cite{OJNS2005,SRK2006}, and in models of growing networks \cite{GL2008}.

The mathematical theory of records has been an important subject of research since the paper by Chandler in 1952 (see, e.g., \cite{Chandler1952,FS1954,Nevzorov1987,ABN1998,SZ1999,Nevzorov2001}). Record statistics is now well understood in the case when the entries $x_i$'s are independent and identically distributed (i.i.d.) random variables. However, in most realistic situations the entries of the time series are correlated and the theory in this case is still piecemeal. For weak correlations, with a finite correlation time, one expects the record statistics for a large $n$ series to be asymptotically similar to the uncorrelated case, but this is no longer true when there are strong correlations between the entries. {One of the simplest and most natural time series with strongly correlated entries corresponds to the one of the positions of a one-dimensional random walk, for which many results are known in the mathematics literature, in particular in the context of fluctuation and renewal theory, and can be found in the classical book by Feller \cite{Feller}. In the physics literature, a comprehensive theory of record statistics for random walks with continuous jumps was worked out by Majumdar and Ziff in \cite{MZ2008} (see also ~\cite{Leuven_Review}).}
Considering a time series defined by
\begin{equation}\label{RW}
x_i=x_{i-1}+\eta_i ,
\end{equation}
with $x_0 =0$ and where the $\eta_i$'s are i.i.d. random variables drawn from a symmetric and continuous jump distribution $f(\eta)$, they showed that the joint probability 
$P(R_n,\tau_1,\tau_2,\cdots ,\tau_{R_n})$ of $R_n$ records in $n$ steps, with respective life-times 
$\tau_i$ ($1\le i\le R_n$), is completely independent of $f(\eta)$ for any $n$, and not just asymptotically for large $n$. This remarkable result, which is a consequence of the so-called Sparre Andersen theorem \cite{SA1954}, includes also L\'evy flights where $f(\eta)\sim 1/\vert\eta\vert^{\mu +1}$ for large $\eta$, with $0<\mu <2$. From the joint probability $P(R_n,\tau_1,\tau_2,\cdots ,\tau_{R_n})$ one can derive
the distribution of the number of records $P(R_n)$ by integrating out the age degrees of
freedom ${\tau_1,\tau_2,\ldots, \tau_{R_n}}$, which thus is also universal for all $n$, i.e., independent of $f(\eta)$. In this paper, we will be mainly interested in $P(R_n)$ and in particular, its first moment, $E(R_n)$, the average number of records up to step $n$. For a symmetric and continuous jump distribution, Majumdar and Ziff found that these quantities are given by the universal formulas, valid for any $n$,
\begin{equation}\label{PR_n-MZ}
P(R_n=m)=\binom{2n-m+1}{n}\, 2^{-2n+m-1},
\end{equation}
and
\begin{equation}\label{ER_n_binom}
E(R_n)= (2n+1)\binom{2n}{n}\, 2^{-2n} \, .
\end{equation}
Note that the expressions in Eqs.\ (\ref{PR_n-MZ}) and\ (\ref{ER_n_binom}) corresponds to taking $R_0=1$ (the initial position is counted as the first record). In particular, for large $n$, Eq.\ (\ref{ER_n_binom}) gives
\begin{equation}\label{ER_n-MZ}
E(R_n)= \sqrt{\frac{4\, n}{\pi}}+O\left(\frac{1}{\sqrt{n}}\right)\ \ \ \ \ (n\to +\infty) \,, 
\end{equation}
independently of the jump distribution $f(\eta)$. Similarly, in the limit $m,\, n\to +\infty$ with fixed $m/\sqrt{n}$, the record number distribution $P(R_n=m)$ in Eq. (\ref{PR_n-MZ}) approaches a scaling form,
\begin{equation}\label{dist_mz.1}
P(R_n=m)\simeq  \sqrt{\frac{\pi}{4\,n}}\, g\left(m \, \sqrt{\frac{\pi}{4\,n}}\right)\, ; 
\quad {\rm with}\quad g(x)= \frac{2}{\pi}\, e^{-x^2/\pi}\, , \quad x\ge 0\,  ,
\end{equation}
where the scaling function $g(x)$ is a half-Gaussian, normalized to unity. As long as $f(\eta)$ is symmetric and continuous, the scaling form in Eq.\ (\ref{dist_mz.1}) is also universal, i.e., independent of the jump distribution
$f(\eta)$.
\begin{figure}
\includegraphics[width = 0.7\linewidth]{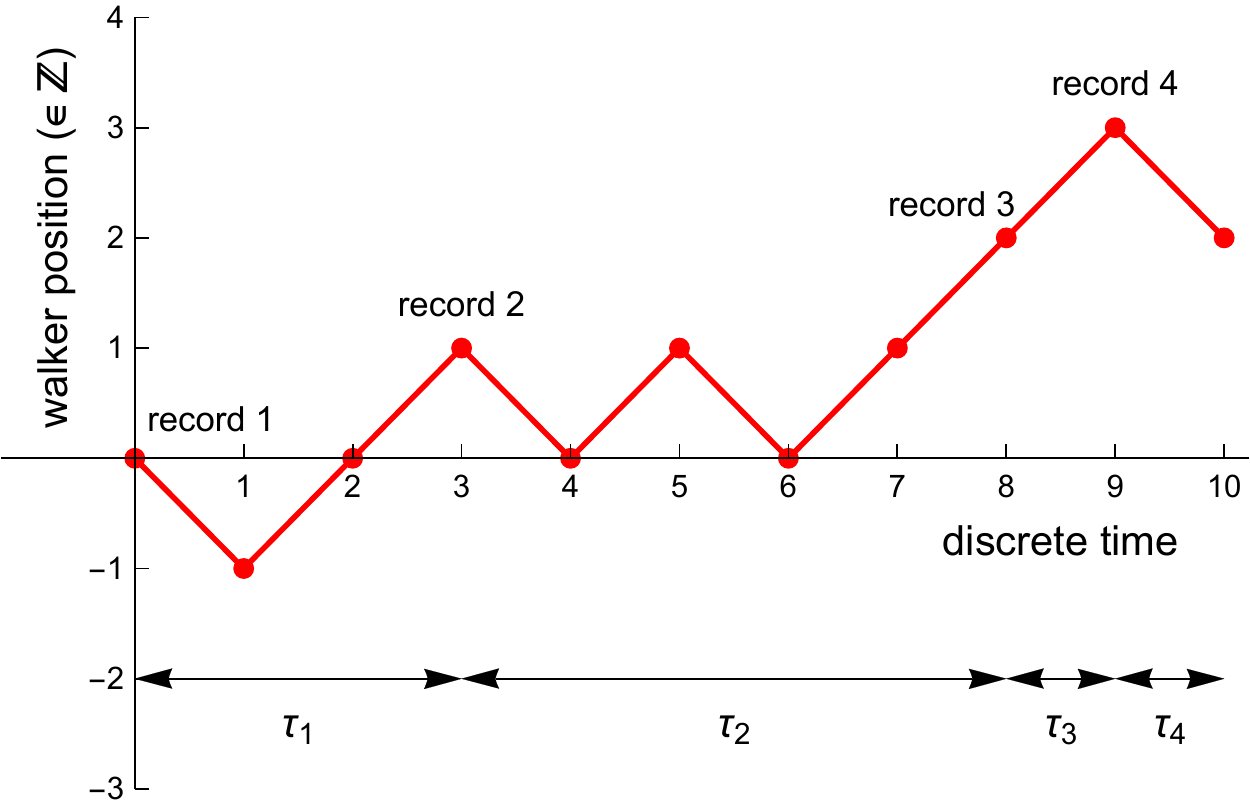}
\caption{Illustration of how we count records for a random walk on a $1D$ lattice. 
A record is counted when the walker position is {\it strictly} greater than all previous positions. 
When the walker position is equal to the one at the last record, it is {\it not} counted as a new record. 
The initial position is counted as the first record. The time interval $\tau_i$ represents the life-time 
of the $i$th record.}\label{figure1}
\end{figure}

Following \cite{MZ2008}, record statistics for random walks has since been studied in different, 
more general, settings (for a recent review on both i.i.d. and random walk cases see\ \cite{GMS2017}). 
This includes, for instance, continuous-time symmetric random walks \cite{Sabhapandit2011}, discrete-time random walks with a constant drift \cite{LW2009,MSW2012}, an ensemble of $N$ independent symmetric random walks \cite{WMS2012}, random walks in the presence of a measurement error and additive instrumental noises~\cite{EKMB2013}, active run-and-tumble particles~\cite{MLMS2020} {or random walks with correlated jumps \cite{K2020}}. The complete universality for all $n$, found in~\cite{MZ2008} for symmetric and continuous $f(\eta)$, does not hold in some of these examples, e.g., in the case of the discrete-time random walk in the presence of a constant 
drift~\cite{MSW2012}. Nevertheless, these generalizations usually concern random walks with continuous jumps and it is natural to ask what happens to record statistics for random walks with {\it discrete} jumps, 
or `lattice random walks'. The question was briefly brought up at the very end of~\cite{MZ2008}. There, it was found that in the particular case where the walker jumps by $\eta =\pm 1$ at each time step, with equal probability $1/2$, Eq.\ (\ref{ER_n-MZ}) is replaced with
\begin{eqnarray}
\label{ER_n-MZ-discrete}
E(R_n)&=&\frac{1}{2}\left\lbrack
\frac{(-1)^{n+1}\Gamma(n-1/2){}_2F_1(\frac{3}{2},-n;\frac{3}{2}-n;-1)}{2\sqrt{\pi}\, \Gamma(n+1)}
+1\right\rbrack \nonumber \\
&=&\sqrt{\frac{2n}{\pi}}+\frac{1}{2}+O\left(\frac{1}{\sqrt{n}}\right)
\ \ \ \ \ (n\to +\infty),
\end{eqnarray}
where ${}_2F_1$ is the hypergeometric function. This is the only known result so far about record statistics 
for random walks with discrete jumps. It can be seen in Eq.\ (\ref{ER_n-MZ-discrete}) that while the expected number of records still grows asymptotically as $\sqrt{n}$ for large $n$, like in the case of a continuous jump distribution in Eq.\ (\ref{ER_n-MZ}), the prefactor $\sqrt{2/\pi}$ of the leading term differs from the one in the continuous case by a factor $1/\sqrt{2}$ and the next correction term $1/2$ is a constant while in the continuous case it goes to zero as $O\left(n^{-1/2}\right)$.

In this paper, we consider a generalized lattice random walk model where the walker stays on a lattice,
but can jump in one step by an arbitrary number of lattice units. Our model includes the $\pm 1$ random walk model as a special case. For this generalized model, we derive the distribution $P(R_n)$
and its first moment $E(R_n)$ explicitly and discuss their asymptotic behavior for large $n$. Our
principal motivation for studying this model is to test how far the universality of record statistics
for symmetric and continuous jump distribution found in Ref. \cite{MZ2008} can be pushed, and
how the discreteness of the position of the walker may modify the record statistics.

The rest of the paper is organized as follows. In section\ \ref{model}, we define our model and summarize our main results. In section\ \ref{sec2}, we determine the generating function of $P(R_n)$ by following a renewal approach, we prove the generalized Sparre Andersen theorem\ (\ref{generalizedSA-intro}), and we use the obtained results to derive the universal scaling form of $P(R_n)$ in the large $n$ limit. In section\ \ref{sec3}, we determine the generating function of the average number of records $E(R_n)$. In the simple case $\eta=\pm 1$, with equal probability $1/2$, we check that this generating function and the associated expression for $E(R_n)$ coincide with the results in\ \cite{MZ2008}. Then, we derive the large 
$n$ asymptotic behavior of $E(R_n)$ in the general case, and we apply the result to random walks with $\eta=0,\, \pm 1,\, \pm 2$, where explicit expressions of the coefficients can be obtained. A scaling form for the large $n$ behavior of $E(R_n)$ is derived in the general case when the probability of staying in place ($\eta =0$) is close to $1$. Section\ \ref{sec4} is devoted to the verification of our analytical predictions via numerical simulations. Finally, we conclude in section\ \ref{sec5}. Some more technical points are relegated to the appendices.
%
%
\section{The model and summary of the main results}\label{model}
We consider a $1D$ random walk starting at the origin, $x_0 =0$, and evolving according to 
Eq. (\ref{RW}) with the jump distribution $f(\eta)$ of the form
\begin{equation}\label{distributionf}
f(\eta)=\sum_{h\in\mathbb{Z}}P(h)\, \delta(\eta -h\alpha),
\end{equation}
where $\alpha >0$ is a given lattice spacing and $0\le P(h)\le 1$ represents the probability to jump by $h$ lattice units in one step, where $h\in \mathbb{Z}$. We further assume that $P(h)$ is symmetric: $P(h)=P(-h)$. Thus, if the walker starts at the origin, it stays on the lattice with lattice spacing $\alpha$ at
any time. Indeed, the position of the walker after the $i$th step is $x_i=l_i\alpha$ where $l_i\in\mathbb{Z}$ evolves according to $l_i =l_{i-1}+h_i$ with $l_0 =0$ and where the jumps $h_i$ are i.i.d. random variables with values in $\mathbb{Z}$ drawn from the symmetric probability $P(h)$. We assume that the Fourier transform of the jump distribution
\begin{equation}\label{FT.1}
\hat{f}(k)=\int_{-\infty}^{+\infty}{\rm 
e}^{ik\eta}f(\eta)\, d\eta =\sum_{h\in\mathbb{Z}}{\rm e}^{ikh\alpha}P(h)
\end{equation}
has the small $k$ behavior $\hat{f}(k)\sim 1-\vert ak\vert^\mu$ ($k\to 0$), where $a\gtrsim\alpha$ is the characteristic length scale of the jumps and $0<\mu\le 2$ is the L\'evy index. As the discrete Fourier transform of $P(h)$, $\hat{f}(k)$ is a periodic function of $k$, unlike the case of continuous jumps where $\hat{f}(k)$ goes to zero as $k\to +\infty$.
%
%
\subsection*{Summary of main results}
Before entering the details of the calculations, it is useful to summarize our main results. 
\begin{itemize}
\item
We show that the renewal property of random walks that was used to study the record statistics for
symmetric and continuous jump distributions in \cite{MZ2008}, can be extended to the present
model, but with a twist. Like in the continuous case, we show that the generating functions of 
$P(R_n)$ and $E(R_n)$ for lattice random walks are respectively given by
\begin{equation}\label{PR_n-gene-intro}
\sum_{n\ge 0}P(R_n=m)\, s^n =\left\lbrack 1-(1-s)\, Q_{\ge 0}(s)\right\rbrack^{m-1}Q_{\ge 0}(s),
\end{equation}
and
\begin{equation}\label{ER_n-gene-intro}
\sum_{n\ge 0}E(R_n)\, s^n =\frac{1}{(1-s)^2 Q_{\ge 0}(s)},
\end{equation}
where $Q_{\ge 0}(s)= \sum_{n\ge 0}q_{\ge 0}(n)\, s^n $ is the generating function of the survival probability,
$q_{\ge 0}(n)={\rm Prob}\left( x_1\ge 0,\, x_2\ge 0,\cdots ,\, x_n\ge 0\left\vert x_0=0\right.\right)$.
The main difference is in the expression for $Q_{\ge 0}(s)$. While in the continuous case, $Q_{\ge 0}(s)= 1/\sqrt{1-s}$ is universal due the Sparre Andersen theorem~\cite{SA1954}, we show that for discrete jump processes on a lattice, it is non-universal and given by
\begin{equation}\label{generalizedSA-intro}
Q_{\ge 0}(s)\equiv\sum_{n\ge 0}q_{\ge 0}(n)\, s^n =\frac{Z(s)}{\sqrt{1-s}},
\end{equation}
where
\begin{equation}\label{zfunction-intro}
Z(s)=\exp\left(-\frac{\alpha}{2\pi}\int_{0}^{\pi/\alpha}
\ln\left\lbrack 1-s\hat{f}(k)\right\rbrack\, dk\right) \, ,
\end{equation}
depends explicitly on the jump distribution $f(\eta)$ through its Fourier transform. The result in 
Eq. (\ref{generalizedSA-intro}) can be derived from a generalized Sparre Andersen theorem~\cite{SA1954} valid for arbitrary jump distributions on a line, discrete or continuous. We also prove this result using an alternative method, which in fact provides a relatively simple non-combinatorial proof of this generalized Sparre Andersen theorem. In the special case of a $\pm 1$ random walk, $f(\eta)= \frac{1}{2}\left[ \delta(\eta- \alpha)+ \delta(\eta+\alpha)\right]$, one has $\hat{f}(k)= \cos(\alpha\, k)$ and $Z(s)$ in Eq. (\ref{zfunction-intro}) can be computed explicitly. One finds $Z(s)= \left[\sqrt{1+s}-\sqrt{1-s}\right]/s$ independent of $\alpha$, thus recovering the known result~\cite{MZ2008,GMS2017}, as it should be. Our formula for $Z(s)$ in Eq. (\ref{zfunction-intro}) generalizes this special case of $\pm 1$ random walks
to arbitrary discrete jump processes on a lattice.
 
\item
Analyzing the general results above for large $n$, we find that while the expected number of records
$E(R_n)$ is non-universal as depending explicitly on $f(\eta)$, remarkably the asymptotic large $n$ behavior of the distribution of the scaled record number $R_n/E(R_n)$ turns out to be universal, i.e.,
independent of $f(\eta)$. More precisely, we show that the distribution $P(R_n=m)$, in the scaling limit $m,\, n\to +\infty$ with $m/E(R_n)$ fixed, takes the scaling form: 
\begin{equation}\label{PR_n-scaling-intro}
P(R_n=m)\sim\frac{1}{E(R_n)}\, g\left(\frac{m}{E(R_n)}\right)\, ; \quad {\rm with} \quad
g(x)= \frac{2}{\pi}\, e^{-x^2/\pi}\, , \quad x\ge 0\, .
\end{equation}
Moreover, the scaling function $g(x)$ coincides rather surprisingly with the one of the symmetric and continuous case in Eq. (\ref{dist_mz.1}). Thus, all the non-universal features are absorbed in $E(R_n)$, while the distribution of the scaled variable $R_n/E(R_n)$ for large $n$ is {\it super-universal}, i.e., 
independent of $f(\eta)$ for both continuous and discrete jump processes.

\item
Finally, our explicit results in Eqs. (\ref{ER_n-gene-intro}), (\ref{generalizedSA-intro}) and 
(\ref{zfunction-intro}) allow us to analyze the large $n$ asymptotic behavior of $E(R_n)$ exactly. We find that for arbitrary discrete jump processes, $E(R_n)$ still behaves as $\sqrt{n}$ for large $n$, like in the continuous case, but now the prefactor and the next subleading term are non-universal as depending explicitly on $f(\eta)$. More specifically, from Eqs.\ (\ref{ER_n-gene-intro}) and\ (\ref{generalizedSA-intro}) near $s=1$, we find that the large $n$ behavior of $E(R_n)$ keeping the terms surviving the $n\to +\infty$ limit only, is given by
\begin{equation}\label{ER_n-asym-intro}
E(R_n)\sim\frac{2}{Z(1)}\, \sqrt{\frac{n}{\pi}} +C_\mu\ \ \ \ \ (n\to +\infty),
\end{equation}
with
\begin{equation}\label{C_mu_intro}
C_2=\frac{\alpha N_{0}}{2\,a\,Z(1)},\ {\rm and}\ C_{0<\mu<2}=0,
\end{equation}
where $N_{0}\ge 1$ is the number of zeros of $1-\hat{f}(k)$ in the interval $0\le k\le\pi/\alpha$. In Section\ 
\ref{sec3d} we give some simple examples where $Z(1)$ in Eqs.\ (\ref{ER_n-asym-intro}) and\ (\ref{C_mu_intro}) can be computed explicitly. For instance, in the simplest case where the walker jumps by $h =0,\, \pm 1$ at each time step, with $P(0)=p$ and $P(\pm 1)=(1-p)/2$ (with $0\le p<1$), Eqs.\ (\ref{ER_n-asym-intro}) and\ (\ref{C_mu_intro}) yield
\begin{equation}\label{heq0pm1-intro}
E(R_n)\sim\sqrt{\frac{2(1-p)\, n}{\pi}}+\frac{1}{2}\ \ \ \ \ (n\to +\infty),
\end{equation}
which shows how the large $n$ behavior in Eq.\ (\ref{ER_n-MZ-discrete}) is modified when the walker is 
allowed to stay in place with probability $p$. The result in Eq. (\ref{heq0pm1-intro}) holds
in the limit $n\to +\infty$ with $0\le p<1$ fixed. It is also interesting to consider the scaling limit when
$p\to 1$, $n\to +\infty$ with the product $n\,(1-p)$ kept fixed. In this scaling limit, we find that
\begin{equation}
E(R_n) \simeq \mathcal{F}\left((1-p)\, n\right) \, , 
\label{ern_scaling.1}
\end{equation}
where the scaling function $\mathcal{F}(x)$ is explicitly given in Eq. (\ref{scalingf-heq0pm1}). We also show in Appendix\ \ref{app2} how to generalize this result to any discrete jump distributions with L\'evy index $0<\mu\le 2$.
\end{itemize}
%
%
\section{Distribution of the number of records: a renewal approach}\label{sec2}

To compute the distribution $P(R_n=m)$ of the number of records $R_n$ after $n$ steps, it turns out to be 
useful to exploit the renewal property of random walks\ \cite{MZ2008} (whether the jumps are continuous or 
discrete). Namely, instead of trying to get the distribution $P(R_n=m)$ of $R_n$ directly (which 
actually seems impossible due to the nontrivial memory carried by the random walk), the strategy would be first to compute the joint probability $P(R_n=m,\, \tau_1,\, \tau_2,\cdots ,\, \tau_m)$ of $R_n$ and the time 
intervals $\tau_1,\, \tau_2,\cdots ,\, \tau_m$ between successive records (see Figure\ \ref{figure1}). Note 
that $\tau_m$ is the time interval between the last record and the arrival time $n$ which does not necessarily correspond to a record. Then, by summing $P(R_n=m,\, \tau_1,\, \tau_2,\cdots ,\, \tau_m)$ over the $\tau_i$'s one can obtain the distribution of records $P(R_n=m)$. In this section, we follow this strategy 
to determine the generating function of $P(R_n=m)$ with respect to $n\ge 0$ which encodes all the information about $P(R_n=m)$.
%
%
\subsection{Generating function of $\bm{P(R_n)}$}
\label{sec2a} 
We define the survival 
probability $q_{\ge 0}(n)$ after $n$ steps by 
\begin{equation}\label{small_q_ge0} q_{\ge 0}(n)={\rm 
Prob}\left( x_1\ge 0,\, x_2\ge 0,\cdots ,\, x_n\ge 0\left\vert x_0=0\right.\right), 
\end{equation} 
for $n\ge 1$ and $q_{\ge 0}(0)=1$. The generating function of $q_{\ge 0}(n)$ is 
\begin{equation}\label{cap_Q_ge0} 
Q_{\ge 0}(s)=\sum_{n\ge 0}q_{\ge 0}(n)\, s^n . \end{equation} 
Note that, since we are considering symmetric random walks and using the translational invariance of the walk with respect to the initial position, one also has 
\begin{eqnarray}\label{small_q_ge0_2} q_{\ge 0}(n)&=&{\rm Prob}\left( x_1\ge x_0,\, x_2\ge x_0,\cdots ,\, 
x_n\ge x_0\left\vert x_0\right.\right) \nonumber \\ &=&{\rm Prob}\left( x_1\le x_0,\, x_2\le x_0,\cdots ,\, 
x_n\le x_0\left\vert x_0\right.\right), 
\end{eqnarray}
independent of the starting point $x_0$. We define the first-crossing time of $x_0$ to be $n$ if
the walker strictly crosses $x_0$ for the first time during the $n$-th step. The first-crossing 
time probability $\varphi(n)$ is then given by 
\begin{equation}\label{phifproba} \varphi(n)= {\rm Prob}\left( 
x_1\le x_0,\, x_2\le x_0,\cdots ,\, x_{n-1}\le x_0,\,  x_n>x_0\left\vert x_0\right.\right), 
\end{equation} 
which is also 
independent of $x_0$ by translation invariance of the walk. From Eqs.\ (\ref{small_q_ge0_2}) and\ 
(\ref{phifproba}), one has the relation 
\begin{equation}\label{phi_small_q_relation} \varphi(n)=q_{\ge 
0}(n-1)-q_{\ge 0}(n), 
\end{equation} 
which implies that the generating function of $\varphi(n)$ can be written in terms of the generating function of $q_{\ge 0}(n)$ as 
\begin{equation}\label{phiproba_gene} 
\Phi(s)\equiv\sum_{n\ge 1}\varphi(n)\, s^n=1-(1-s)Q_{\ge 0}(s). 
\end{equation} 
Now, by decomposing the walk into $m-1$ stretches of duration $\tau_i$ ($1\le i< m$) between the successive records and one final stretch of duration $\tau_m$ between the last record and the last step $n$, and by using the translation invariance with respect to the initial position in each stretch, one finds that the joint probability $P(R_n=m,\, \tau_1,\, \tau_2,\cdots ,\, \tau_m)$ can be written as\ \cite{MZ2008} (see also\ \cite{GMS2017} for further discussion) 
\begin{equation}\label{joint_proba} P(R_n=m,\, \tau_1,\, \tau_2,\cdots ,\, \tau_m)= 
\varphi(\tau_1)\varphi(\tau_2)\cdots\varphi(\tau_{m-1})q_{\ge 0}(\tau_m) \, \delta_{n,\, \sum_{i=1}^{m}\tau_i}, 
\end{equation} 
where the Kronecker delta enforces that the sum of the record ages $\tau_i$ equals the total walk duration $n$. The distribution of records $P(R_n=m)$ is then obtained by summing this expression\  (\ref{joint_proba}) over the $\tau_i$'s, yielding 
\begin{equation}\label{PR_n-renewal} 
P(R_n=m)=\sum_{\tau_1\ge 1}\cdots\sum_{\tau_{m-1}\ge 1}\sum_{\tau_m\ge 0} 
\varphi(\tau_1)\varphi(\tau_2)\cdots\varphi(\tau_{m-1})q_{\ge 0}(\tau_m) \, \delta_{n,\, \sum_{i=1}^{m}\tau_i}. 
\end{equation} 
Taking the generating function of both sides of Eq.\ (\ref{PR_n-renewal}) with respect to $n\ge 0$, one finally obtains 
\begin{equation}\label{PR_n-gene} \sum_{n\ge 0}P(R_n=m)\, s^n 
=\Phi(s)^{m-1}Q_{\ge 0}(s) =\left\lbrack 1-(1-s)\, Q_{\ge 0}(s)\right\rbrack^{m-1}Q_{\ge 0}(s), 
\end{equation} 
which shows that the full statistics of the number of records is entirely encoded in the generating function 
of the survival probability, $Q_{\ge 0}(s)$, that we will now determine.
%
%
\subsection{Generalized Sparre Andersen theorem}\label{sec2b} 
In this subsection, we prove the results in Eqs. (\ref{generalizedSA-intro}) and (\ref{zfunction-intro})
mentioned in section \ref{model}. Below, first we provide a short proof using the generalized Sparre Andersen theorem, and then we give an alternative derivation without assuming this theorem. Our derivation then constitutes a new proof of the generalized Sparre Andersen theorem. 

\vskip 0.4cm

\noindent {\bf Derivation using the generalized Sparre Andersen theorem:} The generalized Sparre Andersen theorem~\cite{SA1954} states that the generating function $Q_{\ge 0}(s)$ of the survival probability is given by the formula, valid for arbitrary jump distributions (discrete or continuous),
\begin{equation}\label{SA_general.1}
Q_{\ge 0}(s)=\sum_{n\ge 0}q_{\ge 0}(n)\, s^n = \exp\left[ \sum_{m=1}^{\infty} \frac{s^m}{m}\, {\rm Prob.}
(x_m \le 0)\right]\, . 
\end{equation}
This is a remarkable theorem as it relates the survival probability $q_{\ge 0}(n)$ up to $n$ steps (which depends on the full history of a trajectory and hence is nonlocal in time) to a local quantity in time, namely
the probability ${\rm Prob.}(x_m \le 0)$ that the walker {\em at} step $m$ is non-positive. Using the symmetry of the walk, it is clear that ${\rm Prob.}(x_m<0)= {\rm Prob.}(x_m>0)$. In addition, the total probability at step $m$ must be unity, i.e., $2 \, {\rm Prob.}(x_m<0)+ {\rm Prob.} (x_m=0)=1$. Hence,
\begin{equation}\label{prob_eq.1}
{\rm Prob.}(x_m\le 0)= {\rm Prob.}(x_m<0)+ {\rm Prob.}(x_m=0)= \frac{1}{2}\left[1+ {\rm Prob.}(x_m=0)\right]\,.
\end{equation}
Substituting\ (\ref{prob_eq.1}) on the right-hand side of Eq. (\ref{SA_general.1}) and using $\sum_{m=1}^{\infty} s^m/m= -\ln (1-s)$ gives
\begin{equation}\label{SA_general.2}
Q_{\ge 0}(s)= \frac{Z(s)}{\sqrt{1-s}}; \quad {\rm where} \quad Z(s)= 
\exp\left[\frac{1}{2}\, \sum_{m=1}^{\infty} \frac{s^m}{m}\, {\rm Prob.}(x_m=0)\right]\, .
\end{equation}
Thus, it remains to compute ${\rm Prob.}(x_m=0)$, which can be done easily. For arbitrary discrete jump processes, it is clear that the position of the walker after $m$ steps, starting from the origin, is simply $x_m= \sum_{i=1}^m \eta_i$, where $\eta_i$'s are i.i.d random variables each drawn from $f(\eta)$. Hence the Fourier transform of ${\rm Prob.}(x_m)$ is given by $E\left({\rm e}^{i\, k\, x_m}\right) = [\hat{f}(k)]^m$. Inverting this Fourier transform and putting $x_m=0$, one obtains
\begin{equation}\label{pxm.0}
{\rm Prob.}(x_m=0)= \frac{\alpha}{2\,\pi}\int_{-\pi/\alpha}^{\pi/\alpha} [\hat{f}(k)]^m\, dk\, ,
\end{equation}
which can be used in the expression\ (\ref{SA_general.2}) of $Z(s)$ to get
\begin{equation}\label{Zs.1}
Z(s)= \exp\left[-\frac{\alpha}{2\pi}\, \int_0^{\pi/\alpha} \ln \left(1-s\, \hat{f}(k)\right)\, dk\right]\, ,
\end{equation}
where we used $\hat{f}(k)= \hat{f}(-k)$. This completes the proof of the results announced in Eqs.\ (\ref{generalizedSA-intro}) and (\ref{zfunction-intro}), in Section \ref{model}.
 
\vskip 0.4cm

\begin{figure}
\includegraphics[width = 0.7\linewidth]{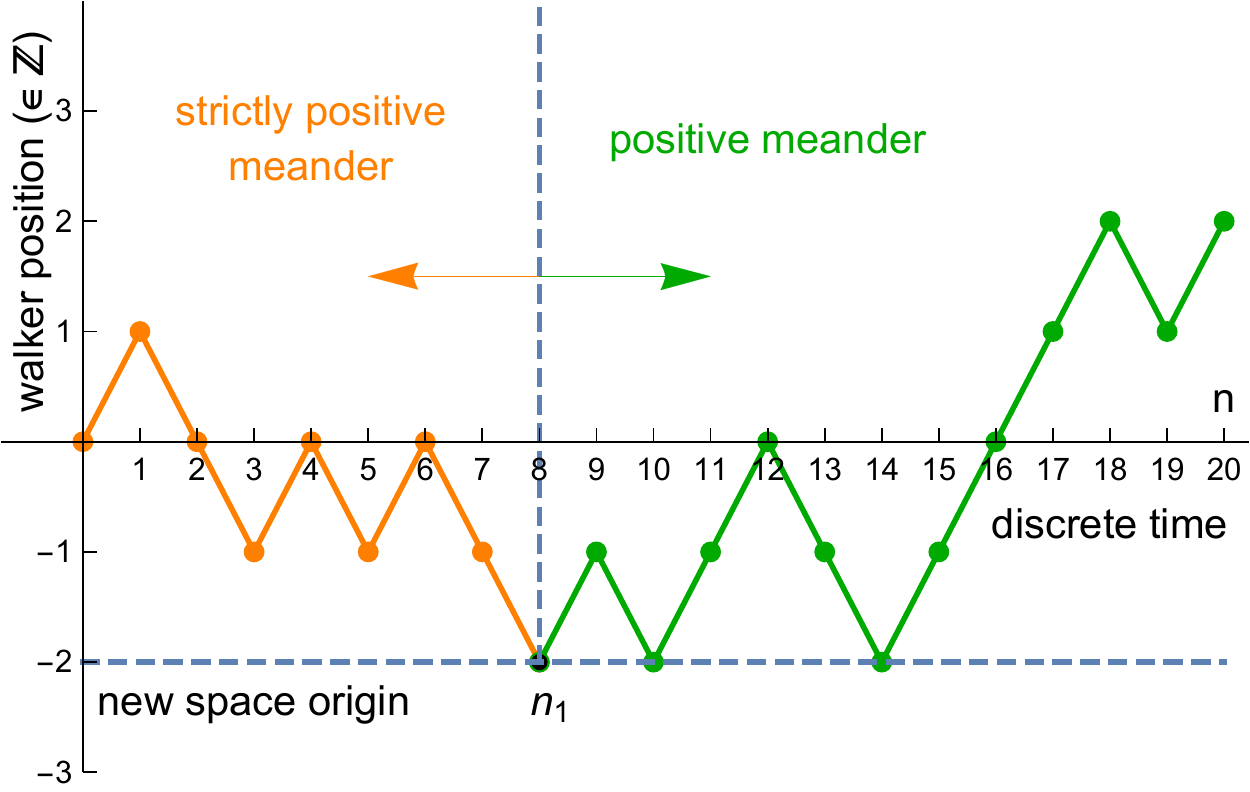}
\caption{A typical trajectory of a discrete jump process of $n$ steps on a $1D$ lattice. We denote
by $n_1$ the step at which the walk reaches for the first time its global minimum within $n$ steps.
The trajectory is split into two intervals, one on the left (shown in orange) and one on the
right (shown in green) of the minimum at $n_1$. Seen from the minimum, the left part of the
trajectory is a {\em strictly} positive meander, since given that the walker arrives for the {\em first time} at
the minimum at $n_1$, it could not have reached this minimum before $n_1$. In contrast, the
right part of the trajectory is just a positive meander, since the walker can reach the same minimum
at some step between $n_1$ and $n$, as illustrated in the figure.}
\label{figure2}
\end{figure}

\noindent {\bf An alternative derivation:} Here we derive the expression of $Q_{\ge 0}(s)$ in Eq.\ (\ref{generalizedSA-intro}) with $Z(s)$ given in Eq.\ (\ref{zfunction-intro}) by an alternative method which does not rely on the generalized Sparre Andersen theorem in Eq. (\ref{SA_general.1}). In fact, the theorem in Eq. (\ref{SA_general.1}) was originally proved by Sparre Andersen using combinatorial arguments~\cite{SA1954}. Our method outlined below actually provides an alternative algebraic/non-combinatorial derivation of this theorem. We proceed in two steps. First, we write 
\begin{equation}\label{small_q_grt0} q_{>0}(n)={\rm Prob}\left( x_1>0,\, x_2>0,\cdots ,\, x_n>0\left\vert 
x_0=0\right.\right), 
\end{equation} 
for $n\ge 1$ and $q_{>0}(0)=1$. The generating function of $q_{>0}(n)$ is 
\begin{equation}\label{cap_Q_grt0} Q_{>0}(s)=\sum_{n\ge 0}q_{>0}(n)\, s^n . 
\end{equation} 
The first step consists in finding the relation between $Q_{\ge 0}(s)$ and $Q_{>0}(s)$. Let $0\le n_1\le n$ denote the step at which the walk reaches its minimum position for the first time and $\pi(n_1)$ the probability of $n_1$. Split the walk into one part from $i=0$ to $i=n_1$ and a second part from $i=n_1$ to $i=n$ ($i$ is the step number) (see Fig. \ref{figure2} for an illustration). Taking the minimum position as a new space origin and inverting the direction of time in the first part, one gets 
\begin{eqnarray}\label{proba_n1} 
\pi(n_1)&=&q_{>0}(n_1)q_{\ge 0}(n-n_1) \nonumber \\ &=&\sum_{n_2\ge 0}q_{>0}(n_1)q_{\ge 0}(n_2)\delta_{n,\, 
n_1+n_2}, 
\end{eqnarray} 
and $\sum_{n_1\ge 0}\pi(n_1)=1$ reads 
\begin{equation}\label{sumproba_n1} 
\sum_{n_1\ge 0}\sum_{n_2\ge 0}q_{>0}(n_1)q_{\ge 0}(n_2)\delta_{n,\, n_1+n_2} =1. 
\end{equation} 
Taking the 
generating function of both sides of Eq.\ (\ref{sumproba_n1}) with respect to $n\ge 0$ and using the 
definitions in Eqs.\ (\ref{cap_Q_ge0}) and\ (\ref{cap_Q_grt0}), one finds the relation 
\begin{equation}\label{Qrelation} Q_{>0}(s)\, Q_{\ge 0}(s)=\frac{1}{1-s}. 
\end{equation} 
Note that in the 
case of a continuous jump distribution, one has $q_{>0}(n)=q_{\ge 0}(n)$. Thus, $Q_{>0}(s)=Q_{\ge 0}(s)$ and Eq.\ (\ref{Qrelation}) immediately gives the Sparre Andersen result $Q_{\ge 0}(s)=1/\sqrt{1-s}$ used in\ 
\cite{MZ2008}, which is a very simple proof of this result.

\begin{figure}
\includegraphics[width = 0.7\linewidth]{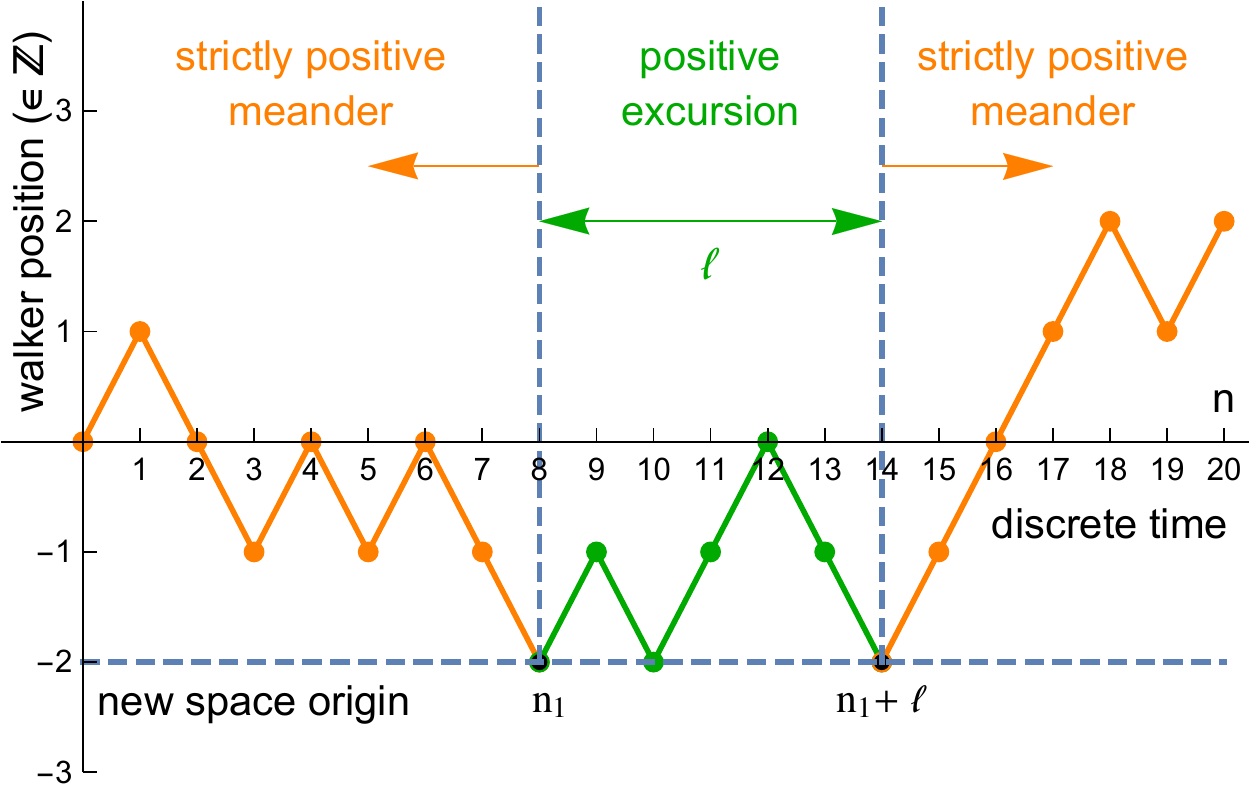}
\caption{A typical trajectory of a discrete jump process of $n$ steps on a $1D$ lattice. We denote
by $n_1$ the step at which the walk reaches for the first time its global minimum within $n$ steps
and by $n_1  + \ell$ the step at which it reaches the global minimum for the last time. 
The parts of the trajectory (shown in orange), on the left of $n_1$ 
and on the right of $n_1 + \ell$, seen from the minimum respectively at $n_1$ and $n_1 + \ell$, are strictly
positive meanders. Between $n_1$ and $n_1 + \ell$, again seen from the minimum at $n_1$, we have a positive (but not necessarily {\em strictly} positive) excursion (shown in green).}
\label{figure3}
\end{figure}

To get the generalized Sparre Andersen theorem for discrete jumps we need to split the walk into three parts, instead of two as we did to get Eq.\ (\ref{Qrelation}). Let $0\le n_1\le n$ denote the step at which the walk reaches its minimum position for the first time, $n_1 +\ell$, with $0\le\ell\le n-n_1$, the step at which it reaches its minimum position for the last time, and $\pi(n_1,\ell)$ the joint probability of $n_1$ and $\ell$ (see Fig. \ref{figure3} for an illustration). We write
\begin{equation}\label{small_b_ge0}
b_{\ge 0}(n)={\rm Prob}\left( x_1\ge 0,\, x_2\ge 0,\cdots ,\, x_n =0\left\vert x_0=0\right.\right),
\end{equation}
for $n\ge 1$ and $b_{\ge 0}(0)=1$. The generating function of $b_{\ge 0}(n)$ is
\begin{equation}\label{cap_B_ge0}
B_{\ge 0}(s)=\sum_{n\ge 0}b_{\ge 0}(n)\, s^n .
\end{equation}
Split the walk into one part from $i=0$ to $i=n_1$, a second part from $i=n_1$ to $i=n_1 +\ell$, and a third part from $i=n_1 +\ell$ to $i=n$. Taking the minimum position as a new space origin and inverting the direction of time in the first part, one gets
\begin{eqnarray}\label{proba_n1andl}
\pi(n_1,\ell)&=&q_{>0}(n_1)b_{\ge 0}(\ell)q_{>0}(n-n_1-\ell) \nonumber \\
&=&\sum_{n_2\ge 0}q_{>0}(n_1)b_{\ge 0}(\ell)q_{>0}(n_2)\delta_{n,\, n_1+n_2+\ell},
\end{eqnarray}
and $\sum_{n_1\ge 0}\sum_{\ell\ge 0}\pi(n_1,\ell)=1$ yields
\begin{equation}\label{sumproba_n1andl}
\sum_{n_1\ge 0}\sum_{n_2\ge 0}\sum_{\ell\ge 0}
q_{>0}(n_1)b_{\ge 0}(\ell)q_{>0}(n_2)\delta_{n,\, n_1+n_2+\ell} =1.
\end{equation}
Taking as previously the generating function of both sides of Eq.\ (\ref{sumproba_n1andl}) with respect to $n\ge 0$ and 
using the definitions in Eqs.\ (\ref{cap_Q_grt0}) and\ (\ref{cap_B_ge0}), one finds
\begin{equation}\label{QBrelation}
Q_{>0}(s)^2\, B_{\ge 0}(s)=\frac{1}{1-s}.
\end{equation}
Defining then $Z(s)=\sqrt{B_{\ge 0}(s)}$ and using Eq.\ (\ref{Qrelation}) on the left-hand side of Eq.\ 
(\ref{QBrelation}), one obtains the generalized Sparre Andersen theorem
\begin{equation}\label{generalizedSA}
Q_{\ge 0}(s)=\frac{Z(s)}{\sqrt{1-s}}\ ; \quad {\rm with} \quad Z(s)= \sqrt{B_{\ge 0}(s)}\, .
\end{equation}
Note that for continuous jumps, all the terms in Eq.\ (\ref{small_b_ge0}) vanish (because the probability that the walker arrives exactly at $x_n=0$ is zero if the jumps are continuous), and $B_{\ge 0}(s)$ in Eq.~(\ref{cap_B_ge0}) reduces to $B_{\ge 0}(s)=b_{\ge 0}(0)=1$ yielding $Z(s)=1$, independently of the (continuous) jump distribution, as expected. For discrete jumps, the expression of $Z(s)$ does depend on the jump distribution. A simple derivation of this expression is given in Appendix\ \ref{app1}. We indeed find that $Z(s)$ coincides with Eq. (\ref{Zs.1}), namely,
\begin{equation}\label{zfunction}
Z(s)=\sqrt{B_{\ge 0}(s)}= \exp\left(-\frac{\alpha}{2\pi}\int_{0}^{\pi/\alpha}
\ln\left\lbrack 1-s\hat{f}(k)\right\rbrack\, dk\right).
\end{equation}
This completes our alternative derivation of the generalized Sparre Andersen theorem. Note that Eqs.\ (\ref{PR_n-gene}),\ (\ref{generalizedSA}) and\ (\ref{zfunction}) entirely characterize the full 
statistics of the number of records for both discrete and continuous jumps (with $Z(s)=1$ in the latter case).
%
%
\subsection{Universal scaling form of $\bm{P(R_n)}$ in the large $n$ limit}\label{sec2c}
Using the generalized Sparre Andersen theorem\ (\ref{generalizedSA}) in Eq.\ (\ref{PR_n-gene}) and inverting the 
generating function, one obtains the following integral representation of $P(R_n)$,
\begin{equation}\label{PR_n-integral1}
P(R_n=m)=\frac{1}{2i\pi}\, \oint
\frac{\left\lbrack 1-Z(s)\sqrt{1-s}\right\rbrack^{m-1} Z(s)}{s^{n+1}\sqrt{1-s}}\, ds.
\end{equation}
We consider the scaling regime defined by $n,\, m\to +\infty$ with $m/\sqrt{n}=O(1)$. Making the change of variable 
$s=\exp(-p/n)$ on the right-hand side of Eq.\ (\ref{PR_n-integral1}) and using the fact that, in the $n\to +\infty$ 
limit, only the vicinity of $s=1$ contributes to the $s$-integral, one finds that at lowest order in the scaling regime, 
the integral representation\ (\ref{PR_n-integral1}) yields
\begin{equation}\label{PR_n-integral2}
P(R_n=m)\sim\frac{Z(1)}{2i\pi\sqrt{n}}\, \int_{\mathcal{L}}\frac{1}{\sqrt{p}}
\exp\left( -Z(1)\frac{m}{\sqrt{n}}\, \sqrt{p}\right)
{\rm e}^{p}dp\ \ \ \ \ (n,\, m\to +\infty ,\ m/\sqrt{n}=O(1)),
\end{equation}
where $\mathcal{L}$ is a Bromwich contour. Performing the inverse Laplace transform on the right-hand side of Eq.\ 
(\ref{PR_n-integral2}) (see, e.g., Eq.\ 29.3.84 in\ \cite{AS1985}), one obtains
\begin{equation}\label{PR_n-scaling-intermed}
P(R_n=m)\sim\frac{Z(1)}{\sqrt{\pi n}}\, \exp\left\lbrack -\left(\frac{Z(1)m}{2\sqrt{n}}\right)^2\right\rbrack
\ \ \ \ \ (n,\, m\to +\infty ,\ m/\sqrt{n}=O(1)).
\end{equation}
From this expression it is straightforward to compute the average number of records in the scaling regime. One has
\begin{eqnarray}\label{ER_n-leading}
E(R_n)&=&\sum_{m\ge 1}mP(R_n=m)\sim
\sum_{m\ge 1}\frac{Z(1)\, m}{\sqrt{\pi n}}\, \exp\left\lbrack -\left(\frac{Z(1)m}{2\sqrt{n}}\right)^2\right\rbrack
\nonumber \\
&\sim&\frac{4}{Z(1)}\, \sqrt{\frac{n}{\pi}}\, \int_0^{+\infty} x{\rm e}^{-x^2}dx
=\frac{2}{Z(1)}\, \sqrt{\frac{n}{\pi}}\ \ \ \ \ (n\to +\infty),
\end{eqnarray}
where we have made the change of variable $x=Z(1)\, m/(2\sqrt{n})$ and replaced the Riemann sum over $m$ with the 
corresponding integral over $x$. Rewriting then the right-hand side of Eq.\ (\ref{PR_n-scaling-intermed}) in terms of 
$E(R_n)$ in Eq.\ (\ref{ER_n-leading}), one gets the universal scaling form
\begin{equation}\label{PR_n-scaling}
P(R_n=m)\sim\frac{1}{E(R_n)}\, g\left(\frac{m}{E(R_n)}\right)
\ \ \ \ \ (n,\, m\to +\infty ,\ m/\sqrt{n}=O(1)),
\end{equation}
with $g(x)=(2/\pi)\, \exp(-x^2/\pi)$. The full statistics of the number of records in the scaling regime is thus 
entirely characterized (at lowest order) by a single parameter, $E(R_n)$, which depends on the jump distribution 
(through $Z(1)$ in Eq.\ (\ref{ER_n-leading})). The scaling function itself, $g(x)$, is universal, i.e., independent of 
the jump distribution. For continuous jumps, $Z(1)$ reduces to $Z(1)=1$ independently of the (continuous) jump 
distribution and the scaling form\ (\ref{PR_n-scaling}) reduces to the one mentioned in\ \cite{MZ2008} (see Eq.\ (\ref{dist_mz.1})), as it should be.

The next section is devoted to a detailed study of $E(R_n)$. We show in particular that in the large $n$ limit and for 
$\mu =2$, there is a constant correction to the leading behavior in Eq.~(\ref{ER_n-leading}). For $0<\mu <2$ there is no 
such a correction and the leading term on the right-hand side of Eq.\ (\ref{ER_n-leading}) is the only term surviving 
the $n\to +\infty$ limit.
%
%
\section{Average number of records}\label{sec3}
As we have just seen, the average number of records, $E(R_n)$, is an important quantity since 
it is the only non-universal parameter that enters in the full distribution of the
number of records in the large $n$ limit. In this section, we study $E(R_n)$ in more details, going further than the simple leading asymptotic behavior\ (\ref{ER_n-leading}). As we did for $P(R_n)$, our starting point is the generating function of $E(R_n)$ with respect to $n\ge 0$ which encodes all the information about $E(R_n)$.
%
%
\subsection{Generating function of $\bm{E(R_n)}$}\label{sec3a}
Multiplying both sides of Eq. (\ref{PR_n-gene}) by $m$, summing over all $m$ and using the relation
$\sum_{m\ge 0}m\, x^{m-1}= 1/(1-x)^2$, one gets
\begin{equation}\label{ER_n-gene-intermed}
\sum_{n\ge 0}E(R_n)\, s^n =\frac{1}{(1-s)^2 Q_{\ge 0}(s)} \, .
\end{equation}
Replacing then $Q_{\ge 0}(s)$ by the generalized Sparre Andersen result\ (\ref{generalizedSA-intro}), one obtains the 
generating function of $E(R_n)$ as
\begin{equation}\label{ER_n-gene}
\sum_{n\ge 0}E(R_n)\, s^n =\frac{1}{(1-s)^{3/2} Z(s)},
\end{equation}
with $Z(s)$ given in Eq.\ (\ref{zfunction-intro}).

In the case of a continuous jump distribution, $Z(s)=1$ and expanding $1/(1-s)^{3/2}$ in power series of $s$, one gets the universal expression of $E(R_n)$ in Eq.\ (\ref{ER_n-MZ}) (see Eq.\ (8) in\ \cite{MZ2008}). Now, we can go further and use the equations\ (\ref{ER_n-gene}) and\ (\ref{zfunction}) to see how this universal result is affected when the random walk takes place on a lattice instead of on the line.
%
%
\subsection{Average number of records for a Bernoulli random walk ($\bm{h_i=\pm1}$)}\label{sec3b}
It turns out that both $\sum_{n\ge 0}E(R_n)\, s^n$ and $E(R_n)$ can be obtained explicitly when the random walk is a symmetric Bernoulli random walk where the walker jumps by $h=\pm 1$ at each step, with $P(\pm 1)=1/2$. This simple example of lattice random walk was dealt with at the end of\ \cite{MZ2008}. Here, we check that the equations\ (\ref{ER_n-gene}) and\ (\ref{zfunction}) do give the same results as the ones in\ \cite{MZ2008}, as it should be.

As mentioned in Section \ref{model}, in this case $\hat{f}(k)$ reduces to $\hat{f}(k)=\cos (\alpha\, k)$ and the integral on the right-hand side of Eq.\ (\ref{zfunction}) yields the explicit expression
\begin{eqnarray}\label{zfunction-Bernoulli}
Z(s)=\left(\frac{2}{1+\sqrt{1-s^2}}\right)^{1/2}
=\frac{2}{\sqrt{1+s}+\sqrt{1-s}},
\end{eqnarray}
independent of $\alpha$. Using Eq.\ (\ref{zfunction-Bernoulli}) in Eq.\ (\ref{ER_n-gene}), one gets
\begin{equation}\label{ER_n-gene-Bernoulli1}
\sum_{n\ge 0}E(R_n)\, s^n =\frac{\sqrt{1+s}+\sqrt{1-s}}{2(1-s)^{3/2}},
\end{equation}
which coincides with the equation (12) in \cite{MZ2008}. Expanding the right-hand side of\ (\ref{ER_n-gene-Bernoulli1}) 
in power series of $s$, one gets
\begin{equation}\label{ER_n-gene-Bernoulli2}
\sum_{n\ge 0}E(R_n)\, s^n =
\sum_{n\ge 1}\frac{1}{2}\left\lbrack\frac{1}{\Gamma(n+1)}\sum_{j=0}^{n}
{n\choose j}\frac{\Gamma(3/2+j)}{\Gamma(3/2+j-n)}
+1\right\rbrack\, s^n,
\end{equation}
and using the relation (see, e.g., Eq. 15.4.1 in \cite{AS1985})
\begin{equation*}
{}_2F_1\left(\frac{3}{2},-n;\frac{3}{2}-n;-1\right)=(-1)^{n+1}
\frac{2\sqrt{\pi}}{\Gamma(n-1/2)}\, \sum_{j=0}^n{n\choose j}
\frac{\Gamma(3/2+j)}{\Gamma(3/2+j-n)},
\end{equation*}
one obtains the expected expression in the first line of Eq.\ (\ref{ER_n-MZ-discrete}) (see Eq. (13) in \cite{MZ2008}),
\begin{equation}\label{ER_n-Bernoulli}
E(R_n)=\frac{1}{2}\left\lbrack
\frac{(-1)^{n+1}\Gamma(n-1/2){}_2F_1(\frac{3}{2},-n;\frac{3}{2}-n;-1)}{2\sqrt{\pi}\, \Gamma(n+1)}
+1\right\rbrack .
\end{equation}
The large $n$ behavior of Eq.\ (\ref{ER_n-Bernoulli}) is more easily obtained from the behavior of Eq.\ 
(\ref{ER_n-gene-Bernoulli1}) near its dominant singularity at $s=1$. One has
\begin{equation}
\sum_{n\ge 0}E({R}_n)\, s^n =
\frac{1}{\sqrt{2}\, (1-s)^{3/2}}+\frac{1}{2\, (1-s)}+O\left(\frac{1}{\sqrt{1-s}}\right)
\ \ \ \ \ (s\to 1),
\end{equation}
which translates into the large $n$ behavior in the second line of Eq.\ (\ref{ER_n-MZ-discrete}),
\begin{equation}\label{anr-Bernoulli-asym}
E({R}_n)=\sqrt{\frac{2n}{\pi}}+\frac{1}{2}+O\left(\frac{1}{\sqrt{n}}\right)
\ \ \ \ \ (n\to +\infty).
\end{equation}
%
%
%
\subsection{Large $\bm{n}$ behavior of $\bm{E(R_n)}$ in the general case}\label{sec3c}
An equation like Eq.\ (\ref{ER_n-Bernoulli}) is a remarkable result and, in most cases, it is not possible to derive such an explicit expression of $E(R_n)$ for all $n$. Nevertheless, as we will see in this section, one can always get the large $n$ behavior of $E(R_n)$, for any lattice random walks and L\'evy flights with index $0<\mu\le 2$.

Again, the large $n$ behavior of $E(R_n)$ is more easily obtained from the behavior of its generating function in Eq.\ 
(\ref{ER_n-gene}) near its dominant singularity at $s=1$. We thus need to determine the behavior of $Z(s)$ near $s=1$. 
To this end, we rewrite Eq.\ (\ref{zfunction}) as
\begin{eqnarray}\label{zfunction-sto1}
Z(s)&=&Z(1)\, \exp\left(-\frac{\alpha}{2\pi}\int_{0}^{\pi/\alpha}
\ln\left\lbrack 1+(1-s)\hat{F}(k)\right\rbrack\, dk\right) \nonumber \\
&=&Z(1)\, \exp\left(-\frac{\alpha N_{0}}{2\pi}\int_{0}^{K/2}
\ln\left\lbrack 1+(1-s)\hat{F}(k)\right\rbrack\, dk\right),
\end{eqnarray}
with $\hat{F}(k)=\hat{f}(k)/[1-\hat{f}(k)]$ and $N_{0}=2\pi/(\alpha K)$, where $K$ is the period of $\hat{f}(k)$. For $s\to 1$, the correction to $Z(s)=Z(1)$ in Eq.\ (\ref{zfunction-sto1}) depends on the behavior of $\hat{F}(k)$ near the zeros of $1-\hat{f}(k)$, where $\hat{F}(k)$ is large. It can be proved that in the interval $0\le k\le K/2$, there is no other zero than the one at $k=0$. Thus, in the limit $s\to 1$, the correction to $Z(s)=Z(1)$ depends on the behavior of $\hat{F}(k)$ near $k=0$, only. For $\mu =2$, we make the change of variable $k=\sqrt{1-s}\, q/a$ and letting $s\to 1$, one gets
\begin{equation}\label{Zasym_mueq2}
Z(s)\sim Z(1)\, \left(1-\frac{\alpha N_{0}}{2a}\sqrt{1-s}\right)\ \ \ \ \ (s\to 1),
\end{equation}
where we have used the small $k$ behavior $\hat{F}(k)\sim\vert ak\vert^{-2}$ and $\int_0^{+\infty}\ln(1+1/q^2)\, dq=\pi$. Similarly, for $1<\mu <2$, we make the change of variable $k=(1-s)^{1/\mu}\, q/a$ and using the small $k$ behavior $\hat{F}(k)\sim\vert ak\vert^{-\mu}$, one finds
\begin{equation}\label{Zasym_1ltmult2}
Z(s)\sim Z(1)\, \left\lbrack 1-O(1-s)^{1/\mu}\right\rbrack\ \ \ \ \ (s\to 1).
\end{equation}
For $0<\mu <1$, $\hat{F}(k)$ is integrable at $k=0$ and writing $\ln\left\lbrack 1+(1-s)\hat{F}(k)\right\rbrack\simeq 
(1-s)\hat{F}(k)$ on the right-hand side of Eq.\ (\ref{zfunction-sto1}), one gets
\begin{equation}\label{Zasym_0ltmult1}
Z(s)\sim Z(1)\, \left\lbrack 1-O(1-s)\right\rbrack\ \ \ \ \ (s\to 1).
\end{equation}
The case $\mu =1$ needs a slightly more careful treatment. Skipping the details, one finds that the algebraic singularity on the right-hand side of Eq.\ (\ref{Zasym_1ltmult2}) changes to a logarithmic singularity, and one has
\begin{equation}\label{Zasym_mueq1}
Z(s)\sim Z(1)\, \left\lbrack 1-O(1-s)\ln\left(\frac{1}{1-s}\right)\right\rbrack\ \ \ \ \ (s\to 1).
\end{equation}
Putting the asymptotic behavior\ (\ref{Zasym_mueq2}) on the right-hand side of Eq.\ (\ref{ER_n-gene}), one gets
\begin{equation}\label{ER_n-gene-asym_mueq2}
\sum_{n\ge 0}E(R_n)\, s^n \sim\frac{1}{(1-s)^{3/2} Z(1)}+\frac{\alpha N_{0}}{(1-s)\, 2aZ(1)}\ \ \ \ \ (s\to 1),
\end{equation}
which translates into the following large $n$ behavior of $E(R_n)$ for $\mu =2$,
\begin{equation}\label{ER_n-asym_mueq2}
E(R_n)\sim\frac{2}{Z(1)}\, \sqrt{\frac{n}{\pi}}+\frac{\alpha N_{0}}{2aZ(1)}\ \ \ \ \ (n\to +\infty).
\end{equation}
By doing the same with the asymptotic behaviors\ (\ref{Zasym_1ltmult2}) to\ (\ref{Zasym_mueq1}), one obtains the large $n$ behavior of $E(R_n)$ for $0<\mu <2$. Keeping the terms surviving the $n\to +\infty$ limit only, one finds
\begin{equation}\label{ER_n-asym_0ltmult2}
E(R_n)\sim\frac{2}{Z(1)}\, \sqrt{\frac{n}{\pi}}\ \ \ \ \ (n\to +\infty).
\end{equation}
Note that the subleading terms of $E(R_n)$ corresponding to the small corrections on the right-hand side of Eqs.\ (\ref{Zasym_1ltmult2}) to\ (\ref{Zasym_mueq1}) go to zero when $n\to +\infty$. Finally, putting the equations\ (\ref{ER_n-asym_mueq2}) and\ (\ref{ER_n-asym_0ltmult2}) together, one obtains the large $n$ behavior of $E(R_n)$ as
\begin{equation}\label{ER_n-asym}
E(R_n)\sim\frac{2}{Z(1)}\, \sqrt{\frac{n}{\pi}} +C_\mu\ \ \ \ \ (n\to +\infty),
\end{equation}
with
\begin{equation}\label{C_mu}
C_2=\frac{\alpha N_{0}}{2aZ(1)},\ {\rm and}\ C_{0<\mu<2}=0,
\end{equation}
valid for any lattice random walks and L\'evy flights with index $0<\mu\le 2$. Since the equation $\hat{f}(k)=1$ has no other solution than $k=0$ in the interval $0\le k\le K/2$, and since $\hat{f}(k)$ is an even periodic function with period $K$, $N_{0}=2\pi/(\alpha K)$ is equal to the number of zeros of $1-\hat{f}(k)$ in the interval $0\le k\le\pi/\alpha$. The presence of $N_{0}$ in Eq.\ (\ref{C_mu}) is crucial to ensure the consistency of the results when there are several different representations of the same physical situations. For instance, $h=\pm 1$ with $\alpha =2$ and $h=\pm 2$ with $\alpha =1$ correspond to the same random walk. Both have the same $a$ and $Z(1)$, but not the same $\alpha$. Without $N_{0}$ one would get two different values of $C_2$, which would make no sense. It can be checked that $\alpha N_{0}=2$ in both cases, yielding the same value of $C_2$. Similarly, $h=\pm 1$ with $\alpha =1$ and $h=\pm 2$ 
with $\alpha =1$ correspond to a mere rescaling of the same walk, which should not affect the statistics of records. Both have the same $\alpha$ and $Z(1)$, but not the same $a$. Again, it can be checked that $N_{0}$ compensates for the difference (one has $N_{0}/a=\sqrt{2}$ in both cases), yielding the same value of $C_2$.

Note also that in the case of a continuous jump distribution, one has $Z(1)=1$ and $\alpha =0$, and Eq.\ 
(\ref{ER_n-asym}) reduces to the large $n$ behavior in Eq.\ (\ref{ER_n-MZ}), valid for all (continuous) jump 
distribution with $0<\mu\le 2$, as it should be.
%
%
\subsection{Application to random walks with $\bm{h =0,\, \pm 1,\, \pm 2}$}\label{sec3d}
As an application, we now consider the random walk on $\mathbb{Z}$, $\alpha =1$, defined by $P(0)=p_0$, $P(\pm 1)=p_1$, 
$P(\pm 2)=p_2$, and $P(h)=0$ otherwise ($h\ne 0,\, \pm 1,\, \pm 2$), with $p_0+2p_1+2p_2=1$. Writing $p=p_0$ and 
$u=p_2/p_1$, one has $p_1=(1-p)/[2(1+u)]$ and $p_2=u(1-p)/[2(1+u)]$. For this random walk, the Fourier transform of the 
jump distribution, $\hat{f}(k)$, reads
\begin{eqnarray}\label{f_of_k}
\hat{f}(k)&=&p_0+2p_1\cos k +2p_2\cos 2k \nonumber \\
&=&1-2p_1(1-\cos k)-2p_2(1-\cos 2k) \nonumber \\
&=&1-\left(\frac{1-p}{1+u}\right)\, \left\lbrack (1-\cos k)+u(1-\cos 2k)\right\rbrack \nonumber \\
&\sim&1-\frac{(1+4u)(1-p)}{2(1+u)}\,k^2\ \ \ \ \ (k\to 0),
\end{eqnarray}
yielding
\begin{equation}\label{a_of_u}
a=\sqrt{\frac{(1+4u)(1-p)}{2(1+u)}}.
\end{equation}
{From the first line of Eq. (\ref{f_of_k}) it is clear that the period of $\hat{f}(k)$ is $K= 2\, \pi$ if $p_1\ne 0$ and $K=\pi$ if $p_1=0$. Since $\alpha=1$, we then have $N_0= 2\pi/(\alpha\, K)=1+\delta_{p_1,0}$. The large $n$ behavior of $E(R_n)$ is then given by the equation\ (\ref{ER_n-asym}) for $\mu =2$ with $\alpha=1$, $N_0=1+\delta_{p_1,0}$, $a$ given in Eq.\ (\ref{a_of_u}), and
\begin{equation}\label{zfunction_seq1}
Z(1)=\sqrt{\frac{1+u}{1-p}}\, z(u),
\end{equation}
where
\begin{equation}\label{smallz_of_u}
z(u)=\exp\left( -\frac{1}{2\pi}\int_0^{\pi}
\ln\lbrack 1-\cos k+u(1-\cos 2k)\rbrack\, dk\right).
\end{equation}
\subsubsection{Large $n$ behaviors of $E(R_n)$ for fixed $p$ and arbitrary $p_1$ and $p_2$}\label{sec3d1}
It turns out that the function $z(u)$ in Eq.\ (\ref{smallz_of_u}) can be computed explicitly. An easy way to do it is to compute its derivative $z'(u)$ as given by the derivative of the right-hand side of Eq.\ (\ref{smallz_of_u}) with respect to $u$, and then integrate the result from $0$ to $u$ with $z(0)=\sqrt{2}$. For arbitrary $u \geq 0$, one obtains
\begin{equation}\label{z_of_u_expl}
z(u) = \frac{2 \sqrt{2}}{1+\sqrt{1+4u}} \;.
\end{equation}
Therefore, substituting (\ref{zfunction_seq1}), (\ref{z_of_u_expl}), and $N_{0}=1+\delta_{p_1,0}$ in Eqs. 
(\ref{ER_n-asym}) and (\ref{C_mu}) one finds
\begin{equation}\label{u_arbitrary}
E(R_n)\sim\sqrt{n(1-p)} \, A(u) + B(u)(1 + \delta_{p_1,0})\ \ \ \ \ (n\to +\infty),
\end{equation}
with
\begin{equation}\label{ABofu}
A(u) = \frac{1+\sqrt{1+4u}}{\sqrt{2\pi(1+u)}} \;\;\; , \;\;\; B(u) = \frac{1}{4}\left(1 + \frac{1}{\sqrt{1+4u}} \right) \;,
\end{equation}
where we recall that $u = p_2/p_1$.

It is clear that for fixed $p_0 = p$, the statistics of the records for walks with steps $h = 0,\, \pm 1$ (i.e. $p_2=0$) and the one with steps $h = 0,\, \pm 2$ (i.e. $p_1=0$), should be exactly the same, for any value of $n$. This implies that taking $u=0$ or $u=+\infty$ in Eq. (\ref{u_arbitrary}) should give exactly the same result. Indeed, one can check from Eq.~(\ref{ABofu}) that $A(0) = A(+\infty) = \sqrt{2/\pi}$ and $B(u)(1+\delta_{p_1,0})\vert_{u=0,\, +\infty}=1/2$, as it should be. Note that since $B(+\infty)=1/4$, the term $\propto \delta_{p_1,0}$ is crucial, as mentioned in the discussion below Eq.~(\ref{C_mu}). It follows in particular that taking $u<+\infty$ arbitrarily large, one gets $\lim_{u\to +\infty}B(u)(1+\delta_{p_1,0})\ne B(u)(1+\delta_{p_1,0})\vert_{u=+\infty}$, which suggests that the limit $p_1 \to 0$ is singular. This might indicate that the limits $n \to \infty$ and $p_1 \to 0$ do not commute, since one can check that for any finite $n$, $E(R_n)$ is a smooth function of $p_1$. Another interesting consequence of the simple fact that $A(0) = A(+\infty)$ is that, since $A(u)$ is a continuous function of $u$, it admits (at least) one extremum. It turns out that $A(u)$ admits a single maximum at $u=2$ for which $A(2) = (2/3)\sqrt{6/\pi}$. {Note that it is not a priori obvious that (i) the function $A(u)$ is a non-monotonic function of $u$ and (ii) that it has a single maximum.}
%
%
%
\subsubsection{Scaling form of $E(R_n)$ for large $n$ and $n(1-p)=O(1)$: $p_2=0$ and general case}\label{sec3d2}
We note that the asymptotic expression in Eq.\ (\ref{u_arbitrary}) requires $(1-p)\, n\gg 1$ to be valid (not just $n\gg 1$). In the opposite limit $(1-p)\, n\ll 1$ corresponding to letting $p\equiv P(h=0)\to 1$ first, then $n\to +\infty$, the walker gets stuck at the initial position $x_0=0$ and there is only one record (the first one at the initial position), yielding $E(R_n)=1$. To conclude this section, we derive a uniform scaling form of $E(R_n)$ in the large $n$ limit which describes the crossover between Eq.\ (\ref{ER_n-asym}) for $(1-p)\, n\gg 1$ and $E(R_n)\sim 1$ for $(1-p)\, n\ll 1$. For simplicity, here we give the details of the calculations for the simple case $h=0,\, \pm 1$ only. The derivation of the scaling form in the general case is explained in detail in Appendix\ \ref{app2}.

First, we determine the large $n$ behavior of $E(R_n)$ as given by Eq.\ (\ref{ER_n-asym}) for $n\to +\infty$ with fixed $p<1$. The case $h=0,\, \pm 1$ corresponds to taking $p_2=0$ (i.e. $u=0$) in Eq. (\ref{u_arbitrary}). It reads, using $A(0) = \sqrt{2/\pi}$ and $B(0)=1/2$,
\begin{equation}\label{ER_n-asym-heq0pm1}
E(R_n)\sim\sqrt{\frac{2(1-p)\, n}{\pi}}+\frac{1}{2}\ \ \ \ \ (n\to +\infty),
\end{equation}
which shows how the large $n$ behavior in Eq.\ (\ref{ER_n-MZ-discrete}) is modified when the walker is allowed to stay in place with probability $p$.

The scaling form of $E(R_n)$ is obtained by considering the limits $n\to +\infty$ and $p\to 1$, keeping the scaling variable $(1-p)\, n$ fixed. Injecting $\hat{f}(k)=p+(1-p)\cos k$, which corresponds to Eq.\ (\ref{f_of_k}) with $u=0$, into Eq\ (\ref{zfunction}), one finds
\begin{equation}\label{zfunction-heq0pm1}
Z(s)=\frac{2}{\sqrt{(1-s)+2s(1-p)}+\sqrt{1-s}}.
\end{equation}
Using Eq.\ (\ref{zfunction-heq0pm1}) on the right-hand side of Eq.\ (\ref{ER_n-gene}), one obtains
\begin{equation}\label{ER_n-gene-heq0pm1}
\sum_{n\ge 0}E(R_n)\, s^n=\frac{\sqrt{(1-s)+2s(1-p)}+\sqrt{1-s}}{2(1-s)^{3/2}},
\end{equation}
from which one gets the following integral representation for $E(R_n)$,
\begin{equation}\label{ER_n-integral1-heq0pm1}
E(R_n)=\frac{1}{2i\pi}\, \oint\frac{\sqrt{(1-s)+2s(1-p)}+\sqrt{1-s}}{2s^{n+1}(1-s)^{3/2}}\, ds.
\end{equation}
Making the change of variable $s=\exp(-\lambda /n)$ in Eq.\ (\ref{ER_n-integral1-heq0pm1}) and using the fact that, in the $n\to +\infty$ limit, only the vicinity of $s=1$ contributes to the $s$-integral, one finds that at lowest order in the scaling regime, the integral representation of $E(R_n)$ reads
\begin{equation}\label{ER_n-integral2-heq0pm1}
E(R_n)\sim\frac{1}{2i\pi}\, \int_{\mathcal{L}}
\frac{\sqrt{\lambda +2n(1-p)}+\sqrt{\lambda}}{2\lambda^{3/2}}\, {\rm e}^{\lambda} d\lambda
\ \ \ \ \ (n\to +\infty ,\, p\to 1),
\end{equation}
where $\mathcal{L}$ is a Bromwich contour. Performing then the inverse Laplace transform on the right-hand side of Eq.\ (\ref{ER_n-integral2-heq0pm1}), one gets the scaling form
\begin{equation}\label{ER_n-scaling-heq0pm1}
E(R_n)\simeq \mathcal{F}\left[\left(1-p\right)\,n\right]\,,
\end{equation}
valid for $n\to +\infty$, $p\to 1$, and fixed $(1-p)\, n=O(1)$, with the scaling function
\begin{equation}\label{scalingf-heq0pm1}
\mathcal{F}(x)=\left\lbrack\left(\frac{1}{2}+x\right)\, I_0(x)
+x\, I_1(x)\right\rbrack\, {\rm e}^{-x}+\frac{1}{2},
\end{equation}
where $I_{\nu}(x)$ is the modified Bessel function of order $\nu$. From the large and small argument behaviors of $I_{\nu}(x)$, one readily obtains
\begin{equation}\label{scalingf-heq0pm1-limits}
\mathcal{F}(x)\sim\left\lbrace
\begin{array}{ll}
\sqrt{2x/\pi}+1/2&(x\to +\infty) \;,\\
1+x/2&(x\to 0) \;,
\end{array}\right.
\end{equation}
from which it can be checked that the scaling form in Eq.\ (\ref{ER_n-scaling-heq0pm1}) describes the 
crossover between Eq.\ (\ref{ER_n-asym-heq0pm1}) for $(1-p)\, n\gg 1$, and $E(R_n)\sim 1$ in the opposite 
limit $(1-p)\, n\ll 1$. Let us note that a similar scaling limit for $E(R_n)$ was studied recently in 
Ref.~\cite{MVK2019} for time series with i.i.d entries $\{x_i\}$ in the context of the Bernoulli model of 
rainfall precipitation records, where a given day is dry with probability $p$ and wet with 
probability $(1-p)$, and a record is counted only for a wet day. Due to the presence of strong correlations 
between the $x_i$'s in the case of random walks with discrete jumps, our scaling function 
$\mathcal{F}(x)$ in Eq.~(\ref{scalingf-heq0pm1-limits}) is however quite different from that in the 
Bernoulli model of i.i.d variables.

It is actually possible to generalize the scaling form in Eq.\ (\ref{ER_n-scaling-heq0pm1}) to any discrete jump 
distribution with L\'evy index $0<\mu\le 2$. The interested reader will find the details of the calculations in 
Appendix\ \ref{app2}. Writing $Z_0(1)=Z(1)\sqrt{1-p}$ independent of $p$ (for fixed ratios $P(h=m)/P(h=1)$, $m\ge 1$), 
one finds
\begin{equation}\label{ER_n-scalingform-general}
E(R_n)\sim\mathcal{G}[(1-p)\, n],
\end{equation}
valid for $n\to +\infty$ and $p\to 1$ with fixed $(1-p)\, n=O(1)$, where the scaling function $\mathcal{G}(x)$ is given by
\begin{equation}\label{scalingfunction-general}
\mathcal{G}(x)=\frac{\sqrt{x}}{2i\pi\, Z_0(1)}\, \int_{\mathcal{L}}
\exp\left(\frac{\alpha}{2\pi}\int_{0}^{\pi/\alpha}
\ln\left\lbrack 1+\frac{\lambda/x}{1-\hat{f}_0(k)}\right\rbrack\, dk\right)\, 
\frac{{\rm e}^{\lambda}}{\lambda^{3/2}}\, d\lambda .
\end{equation}
In Eq.\ (\ref{scalingfunction-general}), $\hat{f}_0(k)$ denotes the Fourier transform of the jump distribution 
corresponding to $p=0$ with fixed ratios $P(h=m)/P(h=1)$, $m\ge 1$. (Note that $Z_0(1)$ is nothing but $Z(1)$ in Eq.\ (\ref{zfunction}) with $\hat{f}_0(k)$ instead of $\hat{f}(k)$). It can then be checked from the large and small argument behaviors of $\mathcal{G}(x)$ given in Eqs.\ (\ref{scalingfunction-largument}) and\ (\ref{scalingfunction-sargument}) that the scaling form in Eq.\ (\ref{ER_n-scalingform-general}) matches smoothly between Eq.\ (\ref{ER_n-asym}) for $(1-p)\, n\gg 1$ and $E(R_n)\sim 1$ for $(1-p)\, n\ll 1$. This result gives the scaling form of the large $n$ behavior of $E(R_n)$ when $P(h=0)\to 1$ in the general case of any discrete jump distribution with L\'evy index $0<\mu\le 2$.
%
%
\section{Numerical simulations}\label{sec4}
\begin{figure}[t]
\includegraphics[width = \linewidth]{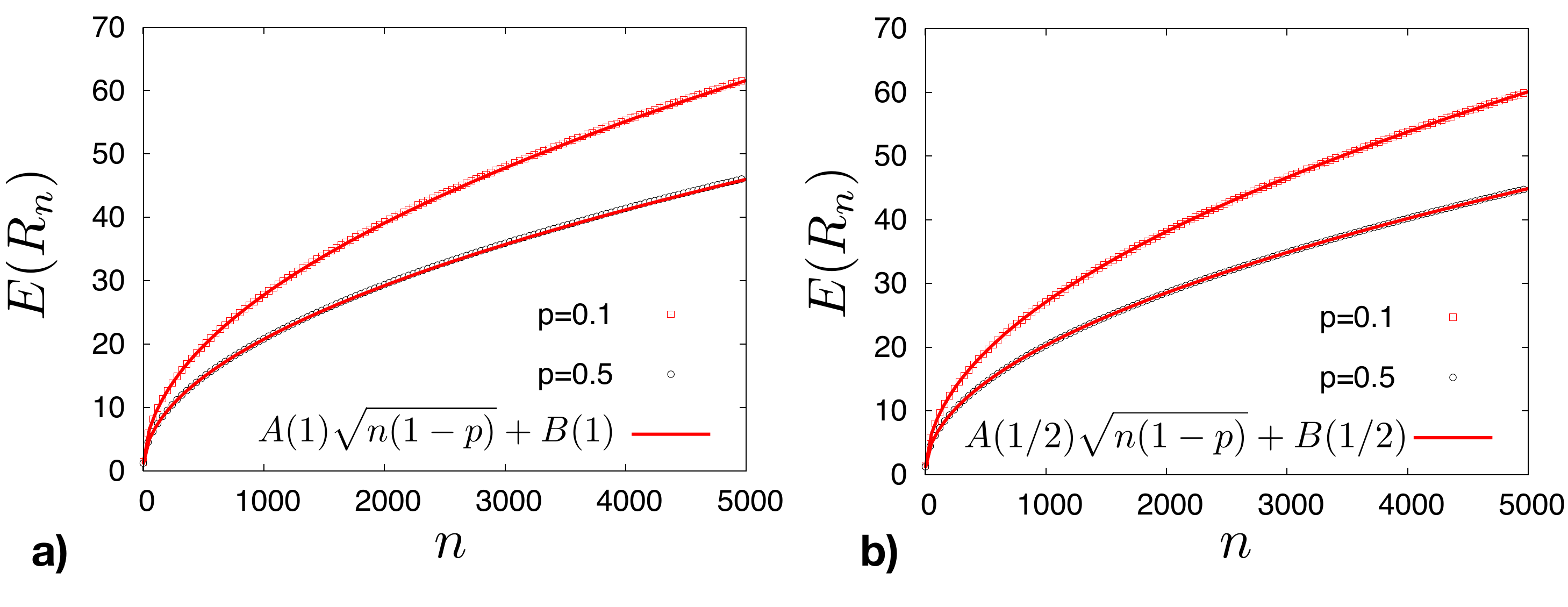}
\caption{Plot of the average number of records $E({R}_n)$ as a function of $n$, for random walks with discrete jumps $h = 0, \pm 1, \pm2$ corresponding to two different values of $u=p_2/p_1$: {\bf a)} $u=1$ and {\bf b)} $u=1/2$. In each panel, the symbols (red squares and black circles) correspond respectively to $p=0.1$ and $p=0.5$ while the solid red line corresponds to the analytical (asymptotic) results given in Eqs. (\ref{u_arbitrary}) and (\ref{ABofu}) with $A(1) =(1+\sqrt{5})/(2\sqrt{\pi})$ and $B(1)=(5+\sqrt{5})/20$ for $u=1$ (left panel), and $A(1/2) = (1+\sqrt{3})/\sqrt{3\pi}$ and $B(1/2)=(3+\sqrt{3})/12$ for $u=1/2$ (right panel). On this scale, we can see that the asymptotic estimate in Eqs.\ (\ref{u_arbitrary}) and (\ref{ABofu}) is very accurate on the whole range of values of $n$.}\label{fig_numerics}
\end{figure}
\begin{figure}[t]
\includegraphics[width = \linewidth]{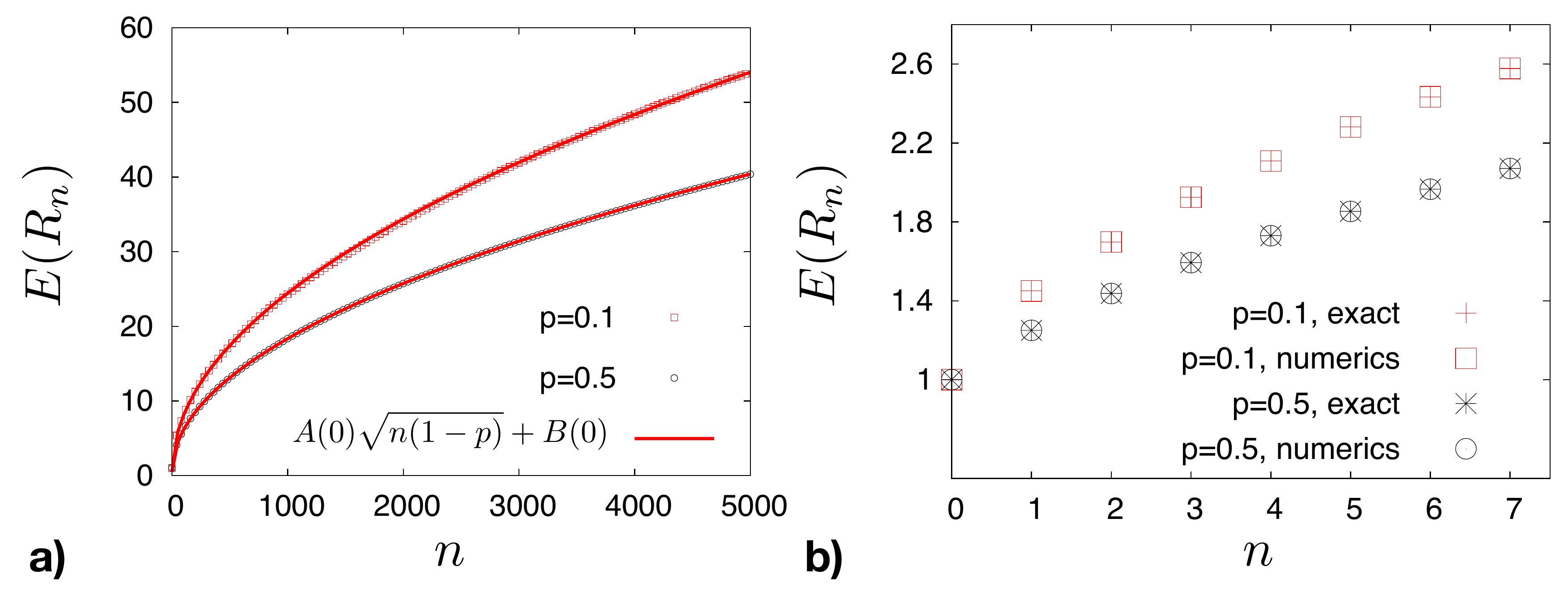}
\caption{Plot of the average number of records $E(R_n)$ as a function of $n$ for random walks with discrete jumps $h = 0, \pm 1$, corresponding to $u=0$. {\bf a)}: The symbols (red squares and black circles) correspond respectively to $p=0.1$ and $p=0.5$ while the solid red line corresponds to the analytical (asymptotic) results given in Eqs. (\ref{u_arbitrary}) and (\ref{ABofu}) with $A(0) = \sqrt{2/\pi}$ and $B(0) = 1/2$. {\bf b)}:~Comparison between our numerical estimates (open symbols as in the left panel) and our exact analytical results (cross symbols) given in Eq. (\ref{exact_rn}) for $p = 0.1$ and $p=0.5$.}\label{fig_numerics_p2eq0}
\end{figure}
In this section, we present numerical simulations of discrete random walks in the case considered in Sec.\ \ref{sec3d}, i.e. $\alpha =1$ and $h=0,\, \pm 1,\, \pm 2$, where exact analytical results can be obtained for the average number of records $E({R}_n)$ in the limit $n \gg 1$ (see Eqs. (\ref{u_arbitrary}) and (\ref{ABofu})). Writing, like in Sec.\ \ref{sec3d}, $p=P(0)$, $p_1 = P(\pm 1)$, and $p_2 = P(\pm 2)$, we study three different cases corresponding to three different values of the ratio $u=p_2/p_1$. Namely, $u=1$, $u=1/2$, and $u=0$. In each case, we have simulated random walks up to $n=5000$ steps for two different values of $p$ (the probability of staying in place): $p=0.1$ and $p=0.5$. The data presented here have been obtained by averaging over $10^5$ independent realizations of the random walk.

In Fig \ref{fig_numerics} a) and b) we show our data for $u=1$ and $u=1/2$, respectively. As we can see, the agreement between the numerics and our analytical (asymptotic) results, given in Eqs.~(\ref{u_arbitrary}) and (\ref{ABofu}), is very good. In Fig. \ref{fig_numerics_p2eq0} we show our data for the case $u=0$, i.e. $p_2 = 0$. Figure\ \ref{fig_numerics_p2eq0} a) shows a comparison between our data for the average number of records $E({R}_n)$ and our asymptotic prediction in Eqs. (\ref{u_arbitrary}) and (\ref{ABofu}) as a function of $n$, with $u=0$, for $p = 0.1$ and $p=0.5$. Here again the agreement is excellent.

Figure \ref{fig_numerics_p2eq0} b) shows a comparison between our numerical computation of $E(R_n)$ and the analytical values obtained from the generating function for $u=0$ and $n=O(1)$. More specifically, using the expression\ (\ref{zfunction}) of $Z(s)$ with $\alpha = 1$ and $\hat f(k) = p + (1-p) \cos k$ (which corresponds to $u=0$) on the right-hand side of Eq.\ (\ref{ER_n-gene}), one gets 
\begin{equation}\label{GF_p0} \sum_{n=0}^\infty E(R_n) s^n = 
\frac{1}{(1-s)^{3/2}} \exp{\left[\frac{1}{2 \pi} \int_0^\pi \ln(1 - sp - s (1-p) \cos k) \right]} \;, 
\end{equation} 
and by expanding the right-hand side in powers of $s$ (which can be done very easily with, e.g., Mathematica) one obtains, for the first values of $n$, 
\begin{eqnarray}\label{exact_rn} &&E(R_0)=1 \;, E(R_1) = \frac{3}{2}-\frac{p}{2} 
\;, \; E(R_2) = \frac{1}{4}(7-2p-p^2) \nonumber \\ &&E(R_3) = 2 - \frac{3p}{4} - \frac{p^3}{4} \;, \; E(R_4) = 
\frac{1}{16}(35-12p-6p^2+4p^3-5p^4) \nonumber \\ &&E(R_5) = \frac{1}{16} (38-15p-10p^3+10p^4-7p^5) \\ &&E(R_6) = 
\frac{1}{32}(81-30p-15p^2+20p^3-45p^4+42p^5-21p^6) \nonumber \\ &&E(R_7) = 
\frac{1}{32}(86-35p-35p^3+70p^4-105p^5+84p^6-33p^7) \nonumber \;. 
\end{eqnarray} 
In Fig. \ref{fig_numerics_p2eq0} b) we compare these exact analytical values with our numerical simulations for $p=0.1$ and $p=0.5$. The comparison shows a perfect agreement.
\begin{figure}[t]
\includegraphics[width = \linewidth]{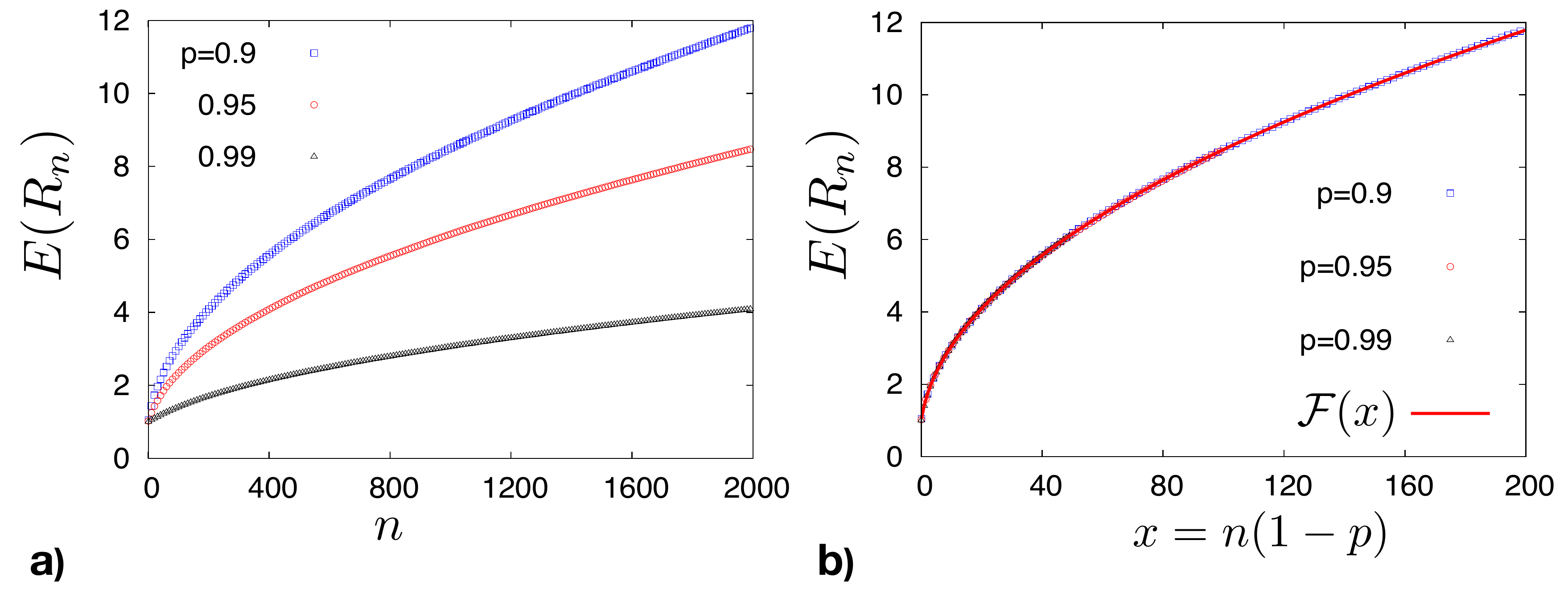}
\caption{{\bf a):} Plots of $E(R_n)$ vs $n$ for three different values of $p$ close to one, namely $p=0.9, 0.95$ and $0.99$ (squares, circles and triangles respectively as indicated in the keys of the main panel). {\bf b)}: Same data as in the left panel but plotted as a function of $x=(1-p)n$, demonstrating a very good collapse of the three curves on a single master cuve, confirming the scaling form predicted in Eq. (\ref{ER_n-scaling-heq0pm1}). The solid red curve is a plot of ${\cal F}(x)$ as given in Eq.~(\ref{scalingf-heq0pm1}).}\label{fig_small_p_numerics}
\end{figure}

Finally, in Fig. \ref{fig_small_p_numerics}, we show plots of $E({R}_n)$ for three different values of $p$ close to one, namely $p=0.9$, $p=0.95$, and $p=0.99$, which confirm the scaling form predicted in 
Eq.~(\ref{ER_n-scaling-heq0pm1}) together with the expression of the scaling function ${\cal F}(x)$ given in Eq.\ (\ref{scalingf-heq0pm1}) -- see Fig. \ref{fig_small_p_numerics} b).
%
%
\section{Conclusion}\label{sec5}
In this paper we have studied a general question concerning the universality of the statistics of the number of records $R_n$ for a discrete-time series whose entries represent the successive positions of an $n$-step random walk in one dimension. Starting at the origin, at each discrete-time step the walker jumps by a certain distance drawn from a symmetric distribution $f(\eta)$. If the jump distribution $f(\eta)$ is continuous, it was known~\cite{MZ2008} that the statistics of records is {\em strongly} universal, i.e., independent of $f(\eta)$ for all $n$. What happens to this universality if the walk takes place on a discrete lattice, i.e., if $f(\eta)$ is nonzero for integer values of $\eta$ only, was still largely unknown. {For the special case of nearest neighbour lattice walk when $\eta=\pm 
1$, previous exact results~\cite{MZ2008} (see also~\cite{GMS2017}) show that while the average number of records $E(R_n)$ still grows asymptotically as 
$\sqrt{n}$ for large $n$, like in the continuous case, the prefactor of this square root growth as well as the next 
subleading term differ from the ones in the continuous case.} This raises the natural question of whether the results from $\pm 1$ walk can be generalized to more general jump processes on a $1D$ lattice, where the walker may jump an arbitrary number of lattice units in one step, without being necessarily restricted to nearest neighbor walk. This question about the record statistics for general discrete jump processes may have relevance, for instance, when the entries of the time series represent the price of a stock on different days. The day-to-day jump in the stock price is often rounded to the nearest integer and stock price effectively moves on a lattice.  The purpose of this paper was to derive the record statistics for such a general discrete jump process on a $1D$ lattice and investigate whether there exist universal results at all, or perhaps the universality holds only asymptotically for large $n$.

Our main conclusion is that the {\em strong} universality, valid for {\em arbitrary} $n$, in the case of continuous jump distributions, no longer holds for discrete jump processes. However, a vestige of universality still remains asymptotically for large $n$. We found that for arbitrary symmetric discrete jump processes, the distribution of the scaled record number $R_n/E(R_n)$ converges, for large $n$, to a universal half-Gaussian distribution, like in the case of continuous jumps. The dependence on the jump distribution $f(\eta)$ is entirely absorbed in the expected number of records $E(R_n)$, leaving the scaled distribution itself independent of $f(\eta)$. By computing the large $n$ behavior of $E(R_n)$ for general discrete jump processes, we found that while it still grows as $\sqrt{n}$ for large $n$ irrespective of $f(\eta)$, the prefactor of $\sqrt{n}$ as well as the next subleading term depend explicitly on the jump distribution.

A technical bonus of our work is an alternative derivation of the celebrated generalized Sparre Andersen theorem that was originally derived using combinatorial arguments~\cite{SA1954}. Our derivation
relies more on an algebraic and non-combinatorial approach. Finally, the techniques developed in this paper will hopefully be useful to study other observables for discrete jump processes on a lattice, such as the statistics of the maximum displacement, the time at which the maximum/minimum occurs etc.

{In this paper, we have restricted ourselves to the study of record statistics for symmetric lattice random walks. A natural question is to ask
how this record statistics will be affected by the presence of a nonzero drift that would make the jump distribution on the lattice asymmetric.  
Indeed, for continuous jump distribution (not necessarily symmetric), the record statistics has been studied using the
generalized Sparre Andersen theorem. In this case, even though the full universality of the record statistics at all step $n$ no longer holds for asymmetric jumps, there remains a vestige of universality in the limit of large $n$ \cite{MSW2012}. It would be interesting to see how the drift affects the record
statistics in the case of lattice random walks studied in this paper.}

%
%
\appendix
\section{Derivation of the expression of $\bm{Z(s)}$}\label{app1}
In this appendix, we derive the expression of the function $Z(s)$ given in Eq.\ (\ref{zfunction}). We write
\begin{equation}\label{small_b_grt0}
b_{>0}(n)={\rm Prob}\left( x_1>0,\, x_2>0,\cdots ,\, x_n =0\left\vert x_0=0\right.\right),
\end{equation}
for $n\ge 1$ and $b_{>0}(0)=1$. The generating function of $b_{>0}(n)$ is
\begin{equation}\label{cap_B_grt0}
B_{>0}(s)=\sum_{n\ge 0}b_{>0}(n)\, s^n .
\end{equation}
Similar expressions for $b_{\ge 0}(n)$ and $B_{\ge 0}(s)$ are given in Eqs.\ (\ref{small_b_ge0}) and\ (\ref{cap_B_ge0}), 
respectively. For $n\ge 1$, one has the relation
\begin{equation}\label{excdecomp}
b_{\ge 0}(n)=\sum_{m=0}^{n-1}b_{\ge 0}(m)b_{>0}(n-m),
\end{equation}
which is easily seen by taking for $m$ the last time at which the walker touches $x=0$ before arriving at $x_n=0$. Writing
$$
\sum_{m=0}^{n-1}b_{\ge 0}(m)b_{>0}(n-m)=
\sum_{m_1\ge 0}\sum_{m_2\ge 1}b_{\ge 0}(m_1)b_{>0}(m_2)\, \delta_{n,m_1+m_2}
$$
and taking the generating function of both sides of Eq.\ (\ref{excdecomp}) with respect to $n\ge 1$ one gets $B_{\ge 
0}(s)-1=B_{\ge 0}(s)(B_{>0}(s)-1)$ which yields
\begin{equation}\label{BBrelation}
B_{\ge 0}(s)=\frac{1}{2-B_{>0}(s)}.
\end{equation}
Now, we consider a bridge of length $n\ge 1$. {Let $n_1$ and $n_2$ respectively denote the first and  last time at which it reaches its minimum before arriving at $x_n=0$. Clearly $0 \leq n_1 \leq n_2 \leq n-1$.} 
For $n_1=0$, the bridge is a positive excursion of 
probability $b_{\ge 0}(n)$ which we write like in Eq.\ (\ref{excdecomp}) with $m=n_2$. For $1\le n_1\le n-1$, we do the 
Vervaat construction~\cite{Vervaat1979} by cutting at $n_1$. One obtains
\begin{eqnarray}\label{bridge1}
{\rm Prob}(x_n=0)&=&\sum_{n_1=0}^{n-1}\sum_{n_2=n_1}^{n-1}
b_{\ge 0}(n_2-n_1)b_{>0}(n-n_2+n_1) \nonumber \\
&=&\sum_{n_1=0}^{n-1}\sum_{\ell =0}^{n-n_1-1}
b_{\ge 0}(\ell)b_{>0}(n-\ell),
\end{eqnarray}
where $\ell =n_2-n_1$. Now, write $\sum_{\ell =0}^{n-n_1-1}=\sum_{\ell =0}^{n}-\sum_{\ell =n-n_1}^{n}$ on the right-hand 
side of Eq.\ (\ref{bridge1}), use the relation
$$
\sum_{\ell =0}^{n}b_{\ge 0}(\ell)b_{>0}(n-\ell)=2b_{\ge 0}(n),
$$
(which follows trivially from Eq.\ (\ref{excdecomp})) in the sum $\sum_{\ell =0}^{n}$, and make the change of variable 
$\ell\to n-\ell$ in the sum $\sum_{\ell =n-n_1}^{n}$. One gets
\begin{eqnarray}\label{bridge2}
{\rm Prob}(x_n=0)&=&\sum_{n_1=0}^{n-1}\left( 2b_{\ge 0}(n)
-\sum_{\ell =0}^{n_1}b_{\ge 0}(n-\ell)b_{>0}(\ell)\right) \nonumber \\
&=&2n\, b_{\ge 0}(n)-\sum_{\ell =0}^{n-1}(n-\ell)\, b_{\ge 0}(n-\ell)b_{>0}(\ell) \nonumber \\
&=&2n\, b_{\ge 0}(n)-\sum_{\ell_1\ge 1}\sum_{\ell_2\ge 0}\ell_1 b_{\ge 0}(\ell_1)b_{>0}(\ell_2)\, 
\delta_{n,\ell_1 +\ell_2}.
\end{eqnarray}
The factor $(n-\ell)$ in the sum over $\ell$ in the second line comes from the fact that the double sum over $\ell$ and 
$n_1$ is over the domain $0\le\ell\le n_1\le n-1$: for each value of $\ell$, $n_1$ takes $(n-\ell)$ values. Taking the 
generating function of both sides of Eq.\ (\ref{bridge2}) with respect to $n\ge 1$, one finds
\begin{equation}\label{bridge_GF}
\sum_{n\ge 1}{\rm Prob}(x_n=0)\, t^n=t\, B^{\prime}_{\ge 0}(t)\left( 2-B_{>0}(t)\right)
=\frac{t\, B^{\prime}_{\ge 0}(t)}{B_{\ge 0}(t)},
\end{equation}
where we have used Eq.\ (\ref{BBrelation}). It remains to divide both sides of Eq.\ (\ref{bridge_GF}) by $t$ and 
integrate over $t$ from $t=0$ to $t=s$, with $B_{\ge 0}(0)=1$. One obtains
\begin{equation}\label{Bexpression}
B_{\ge 0}(s)=\exp\left(\sum_{n\ge 1}{\rm Prob}(x_n=0)\frac{s^n}{n}\right).
\end{equation}
Finally, from $Z(s)=\sqrt{B_{\ge 0}(s)}$, the relation ${\rm 
Prob}\left(x_n=0\right)=(\alpha/2\pi)\int_{-\pi/\alpha}^{\pi/\alpha}\hat{f}(k)^n dk$, and $\hat{f}(k)=\hat{f}(-k)$, one 
obtains
\begin{eqnarray}\label{zfunction-appendix}
Z(s)&=&\exp\left(\frac{1}{2}\sum_{n\ge 1}{\rm Prob}\left(x_n=0\right)
\frac{s^n}{n}\right) \nonumber \\
&=&\exp\left(-\frac{\alpha}{2\pi}\int_{0}^{\pi/\alpha}
\ln\left\lbrack 1-s\hat{f}(k)\right\rbrack\, dk\right).
\end{eqnarray}
%
%
\section{Derivation of the scaling form of $\bm{E(R_n)}$ in the general case}\label{app2}

In this appendix, we derive the scaling form given in Eq.\ (\ref{ER_n-scalingform-general}) for any discrete jump 
distribution with L\'evy index $0<\mu\le 2$. Write $P(h=0)=p$, $P(h=1)=p_1$, and $u_m=P(h=m)/p_1$. From 
$\hat{f}(k)=p+2p_1\sum_{m\ge 1}u_m\, \cos(m\alpha k)$ and $p+2p_1\sum_{m\ge 1}u_m =1$, one gets
\begin{equation}\label{f_of_k-general1}
\hat{f}(k)=1-\frac{(1-p)}{\sum_{\m\ge 1}u_m}\, \sum_{m\ge 1}u_m\lbrack 1-\cos(m\alpha k)\rbrack .
\end{equation}
Introducing $\hat{f}_0(k)$ the Fourier transform of the jump distribution corresponding to $p=0$ with fixed $u_m$ ($m\ge 
1$), we rewrite Eq.\ (\ref{f_of_k-general1}) in the simpler form
\begin{equation}\label{f_of_k-general2}
\hat{f}(k)=1-(1-p)\lbrack 1-\hat{f}_0(k)\rbrack .
\end{equation}
Using the expression\ (\ref{f_of_k-general2}) of $\hat{f}(k)$ on the right-hand side of Eq.\ (\ref{zfunction-sto1}) and 
considering the scaling regime $s,\, p\to 1$ with fixed $(1-p)/(1-s)$, one obtains
\begin{equation}\label{zfunction-scaling-regime}
Z(s)\sim\frac{Z_0(1)}{\sqrt{1-p}}\, 
\exp\left( -\frac{\alpha}{2\pi}\, \int_0^{\pi/\alpha}\ln\left\lbrack 1+
\frac{(1-s)/(1-p)}{1-\hat{f}_0(k)}\right\rbrack\, dk\right),
\end{equation}
where $Z_0(1)$ is given by Eq.\ (\ref{zfunction}) with $s=1$ and $\hat{f}_0(k)$ instead of $\hat{f}(k)$. Now, inverting 
the generating function in Eq.\ (\ref{ER_n-gene}), one gets the integral representation
\begin{equation}\label{ER_n-integral-general}
E(R_n)=\frac{1}{2i\pi}\, \oint\frac{ds}{s^{n+1}(1-s)^{3/2}Z(s)}.
\end{equation}
Making the change of variable $s=\exp(-\lambda /n)$ in Eq.\ (\ref{ER_n-integral-general}), using the fact that, in the 
$n\to +\infty$ limit, only the vicinity of $s=1$ contributes to the $s$-integral, and taking for $Z(s)$ the asymptotic 
expression in Eq.\ (\ref{zfunction-scaling-regime}), one obtains the following scaling form for $E(R_n)$,
\begin{equation}\label{ER_n-scalingform-general-app}
E(R_n)\sim\mathcal{G}[(1-p)\, n],
\end{equation}
valid for $n\to +\infty$, $p\to 1$, and fixed $(1-p)\, n=O(1)$, with the scaling function
\begin{equation}\label{scalingfunction-general-app}
\mathcal{G}(x)=\frac{\sqrt{x}}{2i\pi\, Z_0(1)}\, \int_{\mathcal{L}}
\exp\left(\frac{\alpha}{2\pi}\int_{0}^{\pi/\alpha}
\ln\left\lbrack 1+\frac{\lambda/x}{1-\hat{f}_0(k)}\right\rbrack\, dk\right)\, 
\frac{{\rm e}^{\lambda}}{\lambda^{3/2}}\, d\lambda .
\end{equation}

The large argument behavior of $\mathcal{G}(x)$ is obtained by doing the same analysis as the one for the behavior of 
Eq.\ (\ref{zfunction-sto1}) near $s=1$. Making the substitutions $(1-s)\to \lambda/x$ and $\hat{F}(k)\to 
1/(1-\hat{f}_0(k))$ in the integral on the right-hand side of Eq.\ (\ref{zfunction-sto1}), and using the small $k$ 
behavior $\hat{f}_0(k)\sim 1-(a_0k)^\mu$, with $a_0=a/(1-p)^{1/\mu}$ independent of $p$, one finds
\begin{eqnarray}\label{scalingfunction-largument}
\mathcal{G}(x)&\sim&\frac{\sqrt{x}}{2i\pi\, Z_0(1)}\, 
\int_{\mathcal{L}}\left( 1+\frac{\alpha}{2a_0}\sqrt{\frac{\lambda}{x}}\, \delta_{\mu,\, 2}\right)\, 
\frac{{\rm e}^{\lambda}}{\lambda^{3/2}}\, d\lambda \nonumber \\
&=&\frac{2}{Z_0(1)}\sqrt{\frac{x}{\pi}}+\frac{\alpha}{2a_0Z_0(1)}\, \delta_{\mu,\, 2}\ \ \ \ \ (x\to +\infty),
\end{eqnarray}
where we have kept the terms surviving the $x\to +\infty$ limit, only. Injecting this result onto the right-hand side of 
Eq.\ (\ref{ER_n-scalingform-general-app}) and using the relations $Z_0(1)=Z(1)\sqrt{1-p}$ and $a_0\, Z_0(1)=a\,Z(1)$, 
one can check that the scaling form\ (\ref{ER_n-scalingform-general-app}) coincides with the large $n$ behavior\ 
(\ref{ER_n-asym}) in the limit $(1-p)\, n\gg 1$.

On the other hand, the small argument behavior of $\mathcal{G}(x)$ is obtained from the equation\ 
(\ref{scalingfunction-general-app}) as
\begin{eqnarray}\label{scalingfunction-sargument}
\mathcal{G}(x)&\sim&\frac{\sqrt{x}}{2i\pi\, Z_0(1)}\, \int_{\mathcal{L}}
\exp\left(\frac{\alpha}{2\pi}\int_{0}^{\pi/\alpha}
\ln\left\lbrack\frac{\lambda/x}{1-\hat{f}_0(k)}\right\rbrack\, dk\right)\, 
\frac{{\rm e}^{\lambda}}{\lambda^{3/2}}\, d\lambda \nonumber \\
&=&\frac{1}{Z_0(1)}\exp\left( -\frac{\alpha}{2\pi}\int_{0}^{\pi/\alpha}
\ln\left\lbrack1-\hat{f}_0(k)\right\rbrack\, dk\right)
=1\ \ \ \ \ (x\to 0).
\end{eqnarray}
Using this result on the right-hand side of Eq.\ (\ref{ER_n-scalingform-general-app}), one finds $E(R_n)\sim 1$ in the 
opposite limit $(1-p)\, n\ll 1$. All these results hold for any discrete jump distribution with L\'evy index $0<\mu\le 
2$.

%
%

%
\end{document}